\documentclass[shortbibliography,twocolumn,prl,aps,superscriptaddress,showpacs,amsmath,amssymb,floatfix]{revtex4-1}
\usepackage{graphicx}
\usepackage{amssymb}
\usepackage{amsmath}
\usepackage{epsfig}
\usepackage{color}
\usepackage{mathtools}
\usepackage[colorlinks,linkcolor=blue,anchorcolor=blue,citecolor=blue,urlcolor=blue]{hyperref}
\usepackage{physics}
\usepackage{bm}
\begin{document}

\title{Angular momentum Josephson effect between two isolated condensates}
\author{Wei-Feng Zhuang}
\affiliation{CAS Key Laboratory of Quantum Information, University of Science and Technology of China, Hefei, 230026, People’s Republic of China}
\author{Yue-Xing Huang}
\affiliation{CAS Key Laboratory of Quantum Information, University of Science and Technology of China, Hefei, 230026, People’s Republic of China}
\author{Guang-Can Guo}
\affiliation{CAS Key Laboratory of Quantum Information, University of Science and Technology of China, Hefei, 230026, People’s Republic of China}
\affiliation{Synergetic Innovation Center of Quantum Information and Quantum Physics, University of Science and Technology of China, Hefei, 230026, P.R. China}
\author{Ming Gong}
\email{gongm@ustc.edu.cn}
\affiliation{CAS Key Laboratory of Quantum Information, University of Science and Technology of China, Hefei, 230026, People’s Republic of China}
\affiliation{Synergetic Innovation Center of Quantum Information and Quantum Physics, University of Science and Technology of China, Hefei, 230026, P.R. China}
\date{\today }

\begin{abstract}
We demonstrate that the two degenerate energy levels in spin-orbit coupled trapped Bose gases, coupled by a quenched Zeeman field,  
can be used for angular momentum Josephson effect. 
In a static quenched field, we can realize a Josephson oscillation with period ranging from millisecond to hundreds of milliseconds. Moreover,
by a driven Zeeman field, we realize a new Josephson oscillation, in which the population imbalance may have the same expression as the current in 
the directed current (dc) Josephson effect. When the dynamics of condensate can not follow up the modulation frequency, it enters the self-trapping regime. This new 
dynamics is understood from the time dependent evolution of the constant-energy trajectory in phase space. This model has several salient advantages 
as compared with the previous ones. The condensates are isolated from their excitations by a finite gap, thus can greatly suppress the damping effect induced by 
thermal atoms and Bogoliubov excitations. The oscillation period can be tuned by several order of magnitudes without influencing other parameters. 
In experiments, the dynamics can be mapped out from spin and momentum spaces, thus is not limited by the spatial resolution in absorption imaging. This system can 
serve as a promising platform for of matter wave interferometry  and  quantum metrology. 
\end{abstract}

\maketitle

Josephson effect \cite{josephson1962possible,josephson1974discovery,feynman1965lectures}, as a fundamental phenomenon in quantum mechanics, has been 
widely explored in superconductors \cite{bloch1970josephson, anderson1963probable, buttiker1983josephson}, 
Helium superfluids \cite{sukhatme2001observation, avenel1988josephson}, Bose-Einstein condensate (BEC) \cite{albiez2005direct, levy2007ac, cataliotti2001josephson},
exciton polariton \cite{lagoudakis2010coherent, abbarchi2013macroscopic} and even Fermi gases \cite{valtolina2015josephson}. This phenomenon has been used to construct 
the superconducting quantum interference device (SQUID) \cite{zimmerman1966macroscopic,koch1987quantum,fagaly2006superconducting,ryu2013experimental}, which is sensitive enough in 
measurement of weak quantities, such as magnetic field, temperature and noise, in precision measurement \cite{gallop2003squids, fagaly2006superconducting}. In recent years, this 
device even serves as a basic building block for quantum computation \cite{ofek2016extending, corcoles2015demonstration, kelly2015state}. The 
Josephson effect generally works in two different modes. In the alternating current (ac) mode, a sinusoidal current across the barrier can be observed, with frequency proportional to the bias over it. This mode can be used for terahertz emission \cite{ozyuzer2007emission,tachiki2005emission}. In contrast, in the
directed current (dc) mode, constant current and Shapiro steps \cite{shapiro1963josephson, kleiner1992intrinsic} can be realized by an oscillating bias. 
These novel dynamics were first predicted with ultracold atoms in \cite{smerzi1997quantum, javanainen1986oscillatory} and verified using two weakly coupled condensates \cite{albiez2005direct,shin2005optical, levy2007ac,cataliotti2001josephson} (known as external Josephson effect) and hyperfine levels \cite{zibold2010classical,chang2005coherent} (internal Josephson effect). This nonlinear dynamics can also be realized using momentum \cite{hou2018momentum, bersano2018experimental} and vortices
\cite{gallemi2016coherent}.

This dynamics in ultracold atoms (in a weak trap) is damped from the thermal atoms or Bogoliubov excitations \cite{zapata1998josephson, polo2018damping}. By taking these contributions 
into account, the Josephson oscillator is mapped to a damped pendulum \cite{levy2007ac} or Langevin equation \cite{polo2018damping}, in which 
the damping effect may become significant even when only a small fraction of atoms are excited (for example 20\% at $T = 0.5T_c$) \cite{levy2007ac}. In this 
work, we consider the dynamics of condensates between two isolated condensates realized by spin-orbit coupled Bose gases in a trap, 
in which the degeneracy of the lowest two levels are ensured by time-reversal (TR) symmetry. We investigate the dynamics of the system with a quenched Zeeman field. 
In a constant field, it realizes the ac Josephson effect with period to be tuned in a wide range. By periodically driving this Zeeman field, we realize a new type of 
Josephson oscillation, with population imbalance the same as the current in dc Josephson effect. With the increasing of modulating frequency, it enters the 
self-trapping regime. 
This new dynamics is understood from the time-dependent constant-energy trajectory of the effective Hamiltonian in phase space. This model has some salient advantages as compared with the previous one, from the 
suppressed thermal effect, measurement in spin and momentum space and wide tunability of period. We expect this system to be a useful platform 
for matter wave interferometry \cite{wang2005atom, schumm2005matter} and  quantum metrology \cite{pezze2018quantum}.

\begin{figure}
	\includegraphics[width=0.4\textwidth]{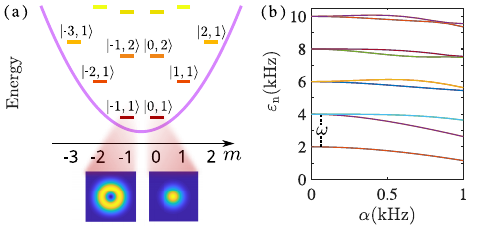}
	\caption{(a) Single particle energy levels labeled by $|m, n\rangle$, where $m$ is the angular 
	momentum and $n$ ($n=1, 2, 3, \cdots$) is the $n$th eigenvalues in ascending order. (b) The single particle spectra as a function 
	of $\alpha$ in a  trap with $\omega = 2 \pi \cdot 318$ Hz. Each line is twofold degenerate from Kramers' degeneracy.}
	\label{fig-fig1}
\end{figure}

\begin{figure}[h!]
    \centering
    \includegraphics[width=0.42\textwidth]{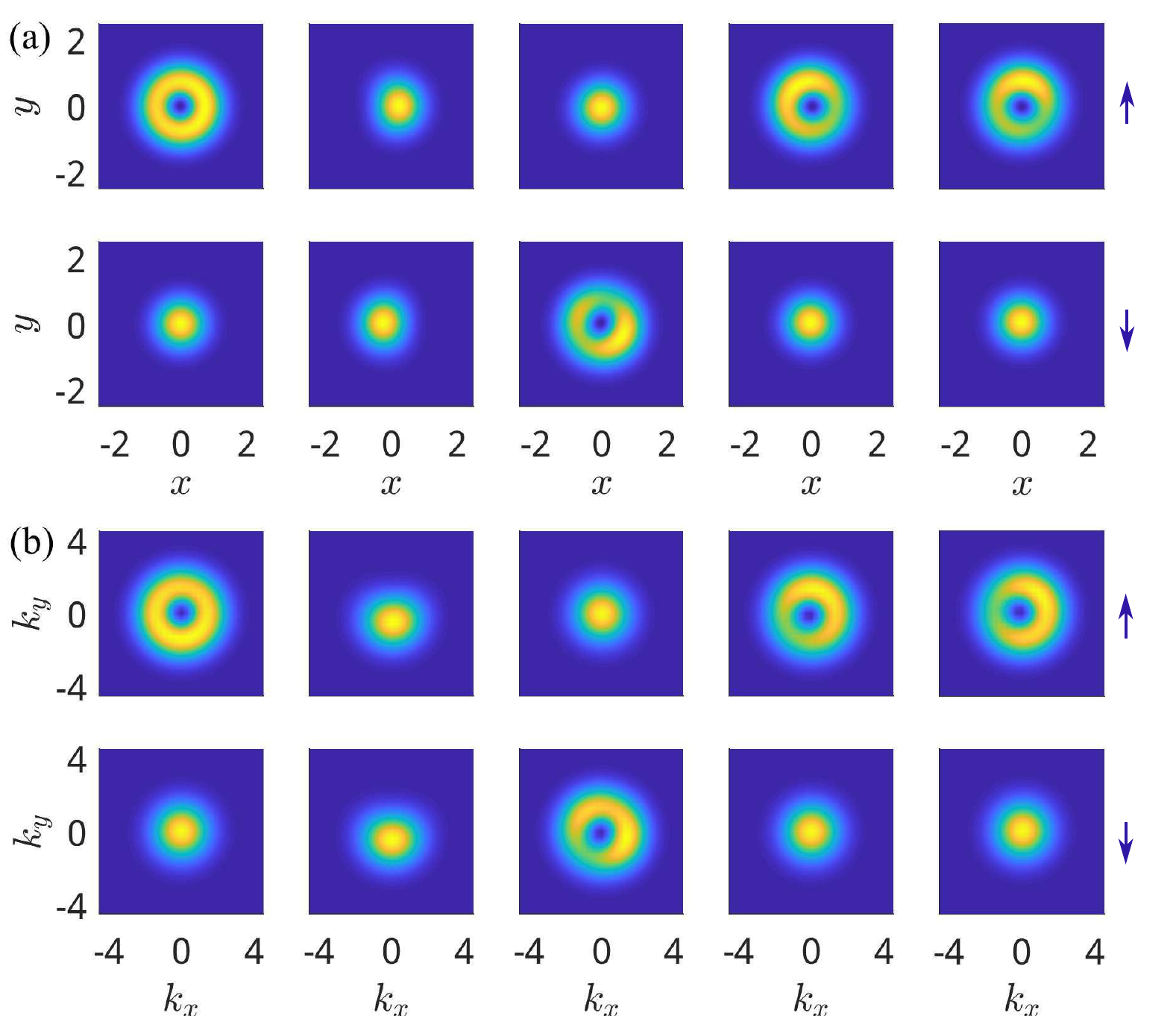}
    \caption{Evolution of the density profile in (top) real space and (bottom) momentum space at 
    $t=0, T/4,  T/2, T, 4T$ with a quenched in-plane Zeeman field. In experiments, 
	this Josephson dynamics can be revealed from the spin space and momentum space.
	Parameters used in simulation are: $\omega = 2\pi \cdot 318$ Hz, $h_x = 2\pi \cdot 80$ Hz, $\alpha_J = 0.9$ and 
	$T =\pi/(\alpha_J h_x)  = 7.3$ ms.}
    \label{fig-fig2}
\end{figure}

We start from the following single particle model in a cylinder trap ($\hbar = 1$) \cite{zhai2015degenerate}, 
\begin{equation}
	\mathcal{H}_0 = {{\bf p}^2 \over 2M} + \alpha (p_x\sigma_y - p_y\sigma_x) + {M \omega^2 {\bf r}^2 \over 2},
\end{equation}
where ${\bf r} = (x, y)$, ${\bf p} = (p_x, p_y)$, $M$ is the mass of the ultracold atom, $\alpha$ is the Rashba spin-orbit coupling (SOC) strength 
\cite{dresselhaus1955spin,bychkov1984oscillatory,liu_effect_2009,lin2009bose,chunji2010spin,lin2011spin,li2013superstripes,zhai2015degenerate,bao2015ground,bhuvaneswari2018spotlighting}, and $\sigma_{x, y}$ are Pauli matrices acting on the hyperfine states labeled by $\sigma = \uparrow, \downarrow$.
In this work, the trapping (angular) frequency should be strong enough, typically of the order of few hundreds Hz, which is comparable with the 
typical temperature in BEC \cite{jochim2003bose, bradley1995evidence}), to suppress the thermal effect from the excited bands. The interaction between the atoms can be written as \cite{wang2010spin,zhai2012spin,zheng2013properties}
\begin{equation}
	\mathcal{V}_\text{I} = \frac{1}{2}\int d^2 \bm r(g_{\uparrow\uparrow} \hat{n}_\uparrow^2+ g_{\downarrow\downarrow}\hat{n}_\downarrow^2+2g_{\uparrow \downarrow} \hat{n}_\uparrow \hat{n}_\downarrow),
\end{equation}
where $\hat{n}_\sigma({\bf r}) = \psi_\sigma^\dagger({\bf r}) \psi_\sigma({\bf r})$. We can define $g_{\sigma\sigma'}=4 \pi a_{\sigma\sigma'}/m$ with 
$a_{\sigma\sigma'}$ being the $s$-wave scattering lengths between the component(s) $\sigma$ and $\sigma'$. Furthermore, we assume $g_{\sigma\sigma}=g$ 
and $g_{\uparrow\downarrow} = g_{12}$ and $c_{12} = g_{12}/g$. The total Hamiltonian for this model can be written as $\mathcal{H} = \mathcal{H}_0 + 
\mathcal{V}_\text{I}$, and the dynamics of each component is determined by the Heisenberg equation $i \dot{\psi}_\sigma = [\psi_\sigma, \mathcal{H}]$.

We can study the single particle spectra in the polar coordinate, then the single particle wave function can be written as
$\mathcal{H}_0 \psi_{m} = \lambda_{m} \psi_{m}$, where \cite{ramachandhran2012half,hu2012spin}, 
\begin{equation}
\psi_m = 
\begin{pmatrix}\psi_{\uparrow,m}(\bm{r})\\
\psi_{\downarrow,m}(\bm{r})
\end{pmatrix}=\frac{1}{\sqrt{2\pi}}\begin{pmatrix}\psi_{\uparrow}(r)\\
\psi_{\downarrow}(r)\text{e}^{i\theta}
\end{pmatrix}\text{e}^{im\theta}.
\end{equation}
We point out that the essential physics here does not relies  on this isotropic symmetry. 
The SOC carries a unit angular momentum, thus couples the two components with different angular momenta $m$ and $m+1$. 
We expand the radial wave function using a two-dimensional harmonic basis 
$R_{k,m}(r)=N_{k,m}\rho^{|m|}\mathcal{L}_{k}^{|m|}(\rho^{2})\text{e}^{-\rho^{2}/2}$, where $k=0,1,2,\cdots$, $m=0,\pm 1, \pm 2, \cdots$, and $\mathcal L_k^{|m|}(\rho^2)$ is the 
Laguerre function, $\rho = r/a_0 $, $a_0=\sqrt{1/(M\omega)}$ is the characteristic length of the harmonic potential and $N_{k,m}=1/a_0 \sqrt{2 k!/(k+|m|)!}$ is the normalized constant. The orthonormal condition is $\int R_{k,m}^{\ast}(\bm{r})R_{k^{\prime},m^{\prime}}(\bm{r})d\bm{r}= \delta_{k,k^{\prime}}\delta_{m,m^{\prime}}$. Then we have 
\begin{eqnarray}
    \psi_{\uparrow}(r) = \sum_{k}a_{k}R_{k,m}(r), \psi_{\downarrow}(r) = \sum_{k}b_{k}R_{k,m+1}(r),
    \label{eq-2}
\end{eqnarray}
and $a_k$ and $b_k$ are determined by the following equation
\begin{equation}
\begin{pmatrix}\mathcal{B} & \mathcal{A}^{\dagger}\\
\mathcal{A} & \mathcal{C}
\end{pmatrix}\begin{pmatrix}a\\
b
\end{pmatrix}=\epsilon\begin{pmatrix}a\\
b
\end{pmatrix},
\end{equation}
where $(a, b)^{T}=(a_{1}, a_{2}, \cdots, a_{k_{c}}, b_{1}, b_{2}, \cdots, b_{k_{c}})^T$,
and $k_{c}$ is the cutoff of radial quantum number $k$, $\mathcal{A}_{k,k^{\prime}}=i\frac{\alpha}{a_{0}}(\sqrt{k+m+1}\delta_{k,k^{\prime}}+\sqrt{k+1}\delta_{k,k^{\prime}+1})$
for $m\ge0$ and $\mathcal{A}_{k,k^{\prime}}=i\frac{\alpha}{a_{0}}(\frac{1}{\sqrt{k-m}}\delta_{k,k^{\prime}}+\frac{1}{\sqrt{(k+1-m)(k-m)}}\delta_{k,k^{\prime}+1})$
for $m<0$, $\mathcal{B}_{k,k^{\prime}} =(2k+|m|+1)\omega \delta_{k,k^{\prime}}$, $\mathcal{C}_{k,k^{\prime}} =(2k+|m+1|+1)\omega \delta_{k,k^{\prime}}$. Without SOC, the two diagonal terms yield the spectra $(2k + |m| +1)\omega$ and $(2k + |m+1| +1)\omega$, with separation $\omega$. 
In Fig. \ref{fig-fig1}, we label these levels as $|m,n\rangle$ by solving the above secular equation, with $n = 1$, $2$, $\cdots$ being the levels in ascending order for a given $m$.

We see that the lowest two states $|-1,1\rangle$ and $|0,1\rangle$ carry angular momenta $m = -1$ and $0$, respectively, with degeneracy in energy 
ensured by TR symmetry $\mathcal{T} = i\sigma_y \mathcal{K}$, where $\mathcal{K}$ is the conjugate operator \cite{cong2011unconventional,hu2012spin}. 
The energy splitting between the ground state and first excited state is roughly determined by the trapping frequency $\omega$ even with SOC (see Fig. \ref{fig-fig1} 
(b)), thus in a typical trap with weak interaction all atoms should be condensed at the lowest two levels. In free space, it is already clear that the 
transition from the plane wave phase to the stripe phase is determined by $c_{12} = 1$, which can be qualitatively changed by the trap potential \cite{hu2012spin}.

The major motivation of this work is to
use these two isolated levels for atomic devices in matter wave interferometry and quantum metrology. We are noticed that these two levels 
could be coupled by an external Zeeman field, which plays the same role as tunneling between two weakly coupled condensates in a double well 
potential \cite{milburn1997quantum,shin2004atom,mustecapliouglu2007quantum, albiez2005direct, levy2007ac, cataliotti2001josephson}. This field may also be used to introduce a chemical potential difference between the two condensates. During the analysis of the dynamics in the quenched Zeeman field in following,
special concern will be payed to what extend various Josephson dynamics can be realized using the experimental accessible parameters.

Let us first consider the dynamics from simulation based on Gross-Pitaevskii equation (GPE), as shown in Fig. \ref{fig-fig2}, which in principle has
taken all bands into account. In this result, we have used a trap frequency $\omega= 2\pi\cdot 318$ Hz, which is comparable to that used in \cite{jin1996collective,mewes1996collective,edwards1996collective, mewes1996bose} and is much smaller than the transverse frequency in one dimensional system \cite{rauer2016cooling,pigneur2018relaxation}.  We focus on the region when initially only one of the state is occupied, which corresponds 
to the physics of plane phase in free space ($c_{12} < 1$). We study the dynamics up on a quenched in-plane 
Zeeman field ($h_x = 2\pi \cdot 80$ Hz). In real space, the initial two components have totally different densities due to different angular momenta. After one full 
period $T = 7.33$ ms, these two density profiles exchange their structure. We find that this dynamics can persist for a long time. It is necessary to point out that 
this period is already much shorter than the Josephson period demonstrated in previous literature, which ranges from 25 to 50 ms based on weakly coupled condensates
\cite{albiez2005direct, levy2007ac, cataliotti2001josephson}.  The similar oscillation can also be found in momentum space, which also oscillates between two different 
density profiles with different angular momenta. This result can be used as smoking gun evidence for angular momentum Josephson effect (see below).

\begin{figure}
  \includegraphics[width=0.43\textwidth]{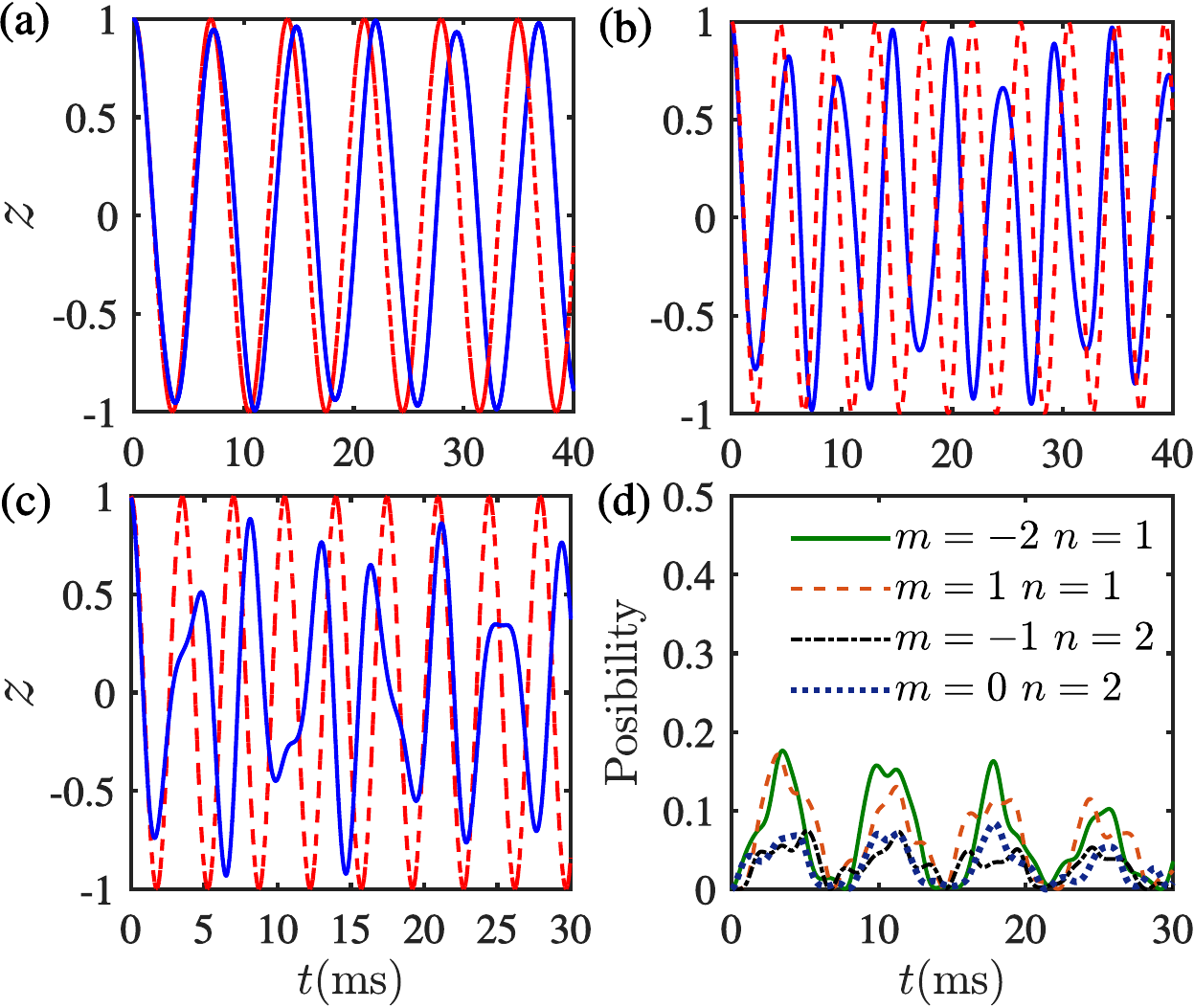}
 \caption{Josephson oscillation with a quenched in-plane Zeeman field in a trap with frequency 
	$\omega = 2\pi\cdot 318$ Hz. The quenched Zeeman fields are $h_x = 2\pi \cdot 80$, $\ 2\pi \cdot 127$, $2\pi \cdot\ 160$ Hz,
	respectively, from (a) to (c). In (a) - (c) the blue solid line and red dashed line are simulated results using GPE and Josephson equation 
	(set $z_0 = 1$). (d) Excited states occupation in (c). Other parameters are: $gN=1200$ Hz, $g_{12}N=1600$ Hz and $\alpha=500$ Hz,
	which yields $\Lambda = V_{12}-2V = 180$ Hz. }
 	\label{fig-fig3}
\end{figure}

We can understand this dynamics in terms of Josephson oscillation. Let us expand the wave function using the lowest two energy levels (two-mode approximation),
\begin{equation}
\psi(\bm r)= \varphi_{-1}(\bm r) a_1+\varphi_0(\bm r) a_2,
\label{eq-13}
\end{equation}
where $a_{1,2}$ are annihilate operators for $|-1, 1\rangle$, and $|0, 1\rangle$, respectively. We obtain an effective Hamiltonian as,
\begin{equation}
H_{\text{eff}}= \sum_{i=1,2}\epsilon_{i}a_{i}^{\dagger}a_{i}+ \sum_i V_i n_i(n_i -1) + V_{12} n_1 n_2,
\end{equation}
where $\varepsilon_1 = \varepsilon_2$ is the single particle energy, $n_i = a_i^\dagger a_i$ and $V_1 = V_2 \equiv V$ is ensured by TR symmetry \cite{V1V2V12}. 
We focus on region $V_{12} > 2V$, thus only one level is occupied. 

\begin{figure}
    \centering
    \includegraphics[width=0.43\textwidth]{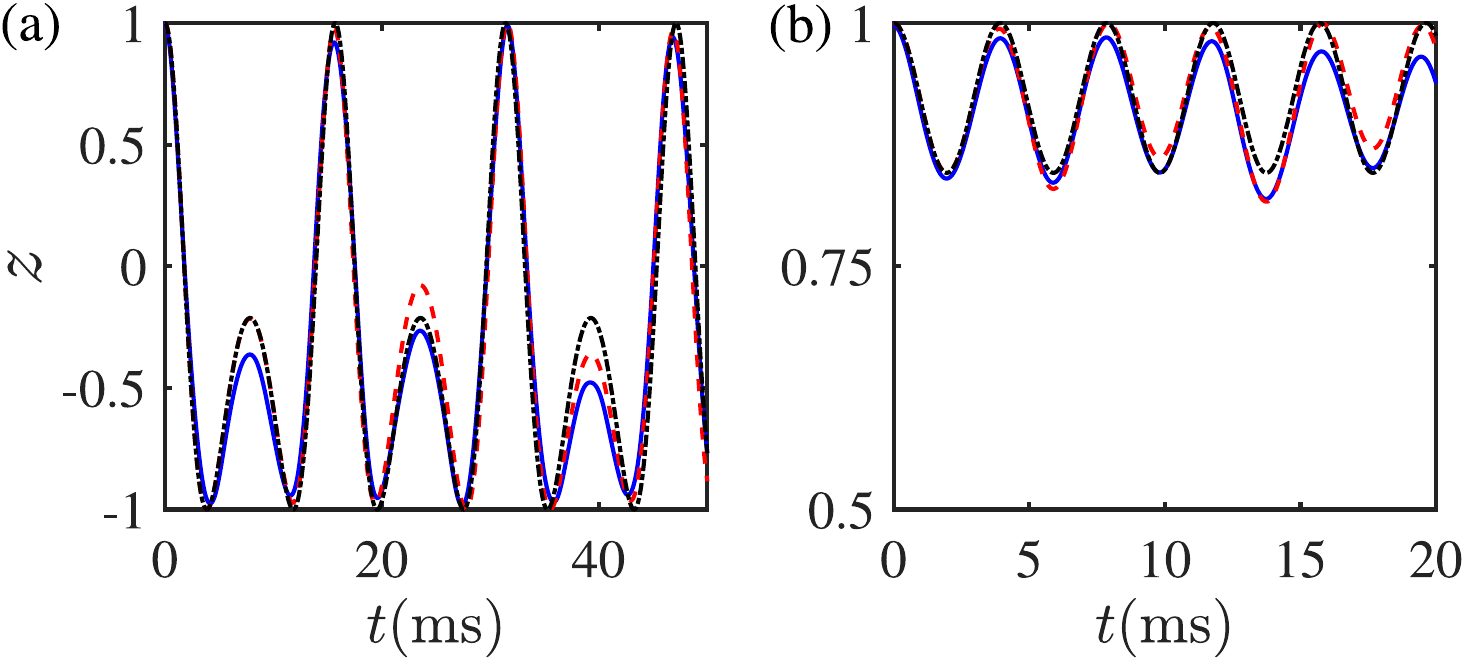}
    \caption{Josephson effect with a driven in-plane Zeeman field. In (a) $h_x=2\pi \cdot 80$ Hz, $\omega_2=2\pi \cdot 32$ Hz; 
	and (b) $h_x = 2\pi \cdot 40$ Hz, $\omega_2 = 2\pi \cdot 127$ Hz. The blue solid line, red dashed line and black dot dashed 
	line are results from GPE, Josephson equation and Eq. \ref{eq-app}, respectively. See Fig. \ref{fig-fig3} for other parameters.}
    \label{fig-fig4}
\end{figure}

We consider the quenched Zeeman field, which couple these two levels by 
\begin{equation}
H_{J}=\int\begin{pmatrix}\varphi_{-1,\uparrow}^{\ast} & \varphi_{-1,\downarrow}^{\ast}\end{pmatrix} h_x \sigma_x 
\begin{pmatrix}
\varphi_{0,\uparrow} \\
\varphi_{0,\downarrow}
\end{pmatrix}d\bm{r} = J a_{1}^{\dagger}a_{2},
\end{equation} 
where $J = h_x \cdot \int\varphi_{\uparrow,-1}^{\ast}(\bm{r})\varphi_{\downarrow,0}(\bm{r})d\bm{r} = \alpha_J h_x$,
with $\alpha_J$ depends strongly on the trap frequency (the phase in $J$ is absorbed into the bosonic operator). 
For $\omega\sim 2\pi \cdot 80 $ Hz, $\alpha_J \sim 0.75$, while for $\omega = 2\pi \cdot 318$ Hz, $\alpha_J \simeq 0.9$. 
The dynamical period is then determined by $T = \pi/\alpha_J h_x$ (see Fig. \ref{fig-fig2}). 
We may also define $\delta \epsilon_m = h_x \langle m| \sigma_x| m\rangle$ as the first-order energy shift, and by TR symmetry, $\delta \epsilon_{-1} 
= -\delta \epsilon_0$. In case of in-plane Zeeman field, $\delta \epsilon_{m} = 0$; while for Zeeman field along $z$ direction, it can be
nonzero. With this quenched field, the dynamics of condensate is given by 
$i\dot{a}_1 = (2V\hat{n}_{1}+V_{12}\hat{n}_{2}) a_1 + J a_2$ and 
$i\dot{a}_2 = (2V\hat{n}_{2}+V_{12}\hat{n}_{1}) a_2 + J a_1$. Then following the particle-phase representation $a_{i}= \sqrt{N_{i}}e^{i\theta_{i}}$ 
, we obtain the following Josephson equation \cite{milburn1997quantum,shin2004atom,mustecapliouglu2007quantum, albiez2005direct}
\begin{equation}
\dot{z}=-2J\sqrt{1-z^{2}}\sin\theta, \quad \dot{\theta}=\Lambda z+\frac{2J z}{\sqrt{1-z^{2}}}\cos\theta,
\label{eq-Josephson effect}
\end{equation}
where $\theta = \theta_2 - \theta_1$ is the phase difference between the condensates, and $z=(N_\uparrow-N_\downarrow)/(N_\uparrow+N_\downarrow)$ 
is the population imbalance and $\Lambda = V_{12} - 2V$. The solution can be well approximated by $z(t) = \cos(2 J t)$ and $\theta \simeq -\pi \text{sign}
(\cos(2J t))$, thus the period is given by $T = \pi/J$.

\begin{figure}[h!]
    \centering
    \includegraphics[width=0.42\textwidth]{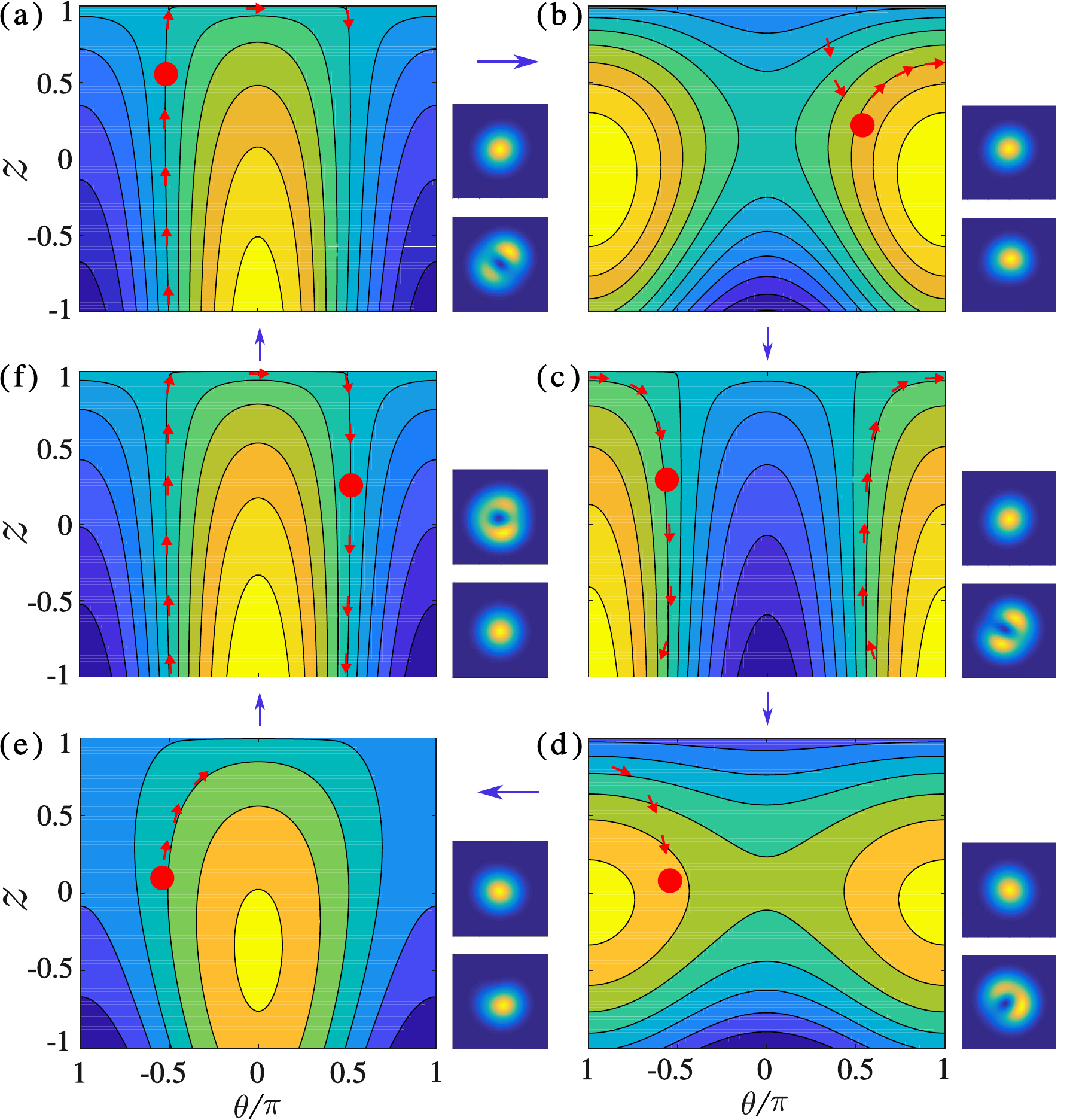}
    \caption{Evolution of the condensate in one full period for Fig. \ref{fig-fig4} (a).  From (a) to (f), $t = $  $0 \sim 8$ ms, 
	$8 \sim 12 $ ms, $12 \sim 19 $ ms, $19 \sim 23$ ms, $23 \sim 25$ ms, and $25 \sim 38$ ms. The inflection points in (b) is $t = 7.8$ ms and 
	(d) is $t = 23.4$ ms. The density profiles in momentum space are given at $t = 4.2$, $7.8$, $11.5$, $19.6$, $23.4$, $31.6$ ms from (a) to (f).}
    \label{fig-fig5}
\end{figure}

We characterize the dynamics using the population imbalance $z$ in Fig. \ref{fig-fig3}. When $h_x \ll \omega$, perfect sinusoidal dynamics by the above
approximated solution can be observed. However, with the increasing of $h_x$, the excited bands are also occupied (see $\eta = h_x/\omega \sim 50\%$ 
in Fig. \ref{fig-fig3} (d)). Empirically we find that when $\eta \le 30\%$, the sinusoidal dynamics can be regarded as excellent within 5\% uncertainty. 
This bound can be further improved by increasing the trap frequency, which suppresses the excited states occupation. For this reason, the period $T$ 
can be tuned in a wide range by controlling the quenched field.

This model can also be used to explore the dc Josephson effect \cite{smerzi1997quantum,albiez2005direct,bloch2005ultracold,liu2007josephson,trombettoni2001discrete}  by applying a modulating chemical potential 
realized with a Zeeman field along $z$ direction. Here we are more 
interested in a new type of nonlinear Josephson effect that has never been explored in previous literature. Let us apply an oscillating in-plane 
Zeeman field by assuming $J = J(t)$ in Eq. \ref{eq-Josephson effect}. This model has an interesting limit, which can be solved exactly. 
In the case of small amplitude oscillation ($|\theta| \ll 1$ and $|z| \ll 1$), we may linearize the above Josephson equations and obtain two new linear 
differential equations: 
$\frac{dz}{dt} =-2J(t) \theta$ and  $\frac{d\theta}{dt} = 2J(t)z$, yielding
\begin{equation}
\begin{cases}
z(t)=z_{0}\cos(\Theta)-\theta_{0}\sin(\Theta),\\
\theta(t)=\theta_{0}\cos(\Theta)-z_{0}\sin(\Theta),
\end{cases}
\label{eq-app}
\end{equation}
where $\Theta =\int_0^t J(t')dt'$ and $z(0) = z_0$ and $\theta(0) = \theta_0$ are the corresponding initial conditions. 
In experiments, these two initial conditions can be completely determined by direct measuring the time dependent population imbalance $z(t)$. 
We are mainly interested in the coupling of the form $J(t) = J_0 + J \cos(\omega_2 t)$, in which case it is interesting to find that the dynamics 
of $z(t)$ is the same as the current ($I \propto \dot{z}$) in the standard dc Josephson effect by modulating of chemical potential 
\cite{DCJ}. As a result, in frequency domain, this dynamics exhibits multiple frequencies as $\pm 2J_0 + n \omega_2$, 
with $n \in \mathbb{Z}$, which can be extracted from the dynamics of $z(t)$ by performing a Fourier transformation. 
Throughout this work, we only consider the simplest case with $J_0 = 0$. The  numerical results are presented  
in Fig. \ref{fig-fig4}. In the case with small oscillation frequency, the dynamics will 
exhibit some strong non-sinusoidal dynamics. With the increasing of $\omega_2$ in region that $J \ll \omega_2 < \omega$ \cite{strongomega2} when 
the dynamics of condensate can not follow up the modulation potential, we realize a self-trapping phase. We need to emphasize that the
mechanism for this self-trapping phase is different from that in previous literature, which is induced by large 
interaction diference $\Lambda$ \cite{smerzi1997quantum, albiez2005direct}. 
These results demonstrate the validity of the approximated solution in Eq. \ref{eq-app} in a much wider range of parameter. More 
intriguing nonlinear dynamics can be realized by engineering of $J(t)$.

We can understand this dynamics from the effective Hamiltonian perspective in phase space, which reads as
\begin{equation}
H= \frac{1}{2}\Lambda(1-z^{2}) + 2J\cos(\omega_2 t)\sqrt{1-z^{2}}\cos\theta.
\end{equation}
During the evolution of $z$ and $\theta$, the constant-energy profile of the Hamiltonian in phase space also changes. The dynamics in a full period for 
Fig. \ref{fig-fig4} (a) is presented in Fig. \ref{fig-fig5}, in which the condensate should evolve almost in the constant-energy trajectory. 
During the evolution of the condensate, the sudden change of constant-energy pattern in the phase space will lead to the inflection points observed
at $t = 7.8$ ms (see Fig. \ref{fig-fig5} (b)) and $t = 23.4$ ms (see Fig. \ref{fig-fig5} (d) - (e)). In this case, let $J = 2 \pi \cdot 71.6$ Hz 
($\alpha_J = 0.9$), and $\omega_2 = 2\pi\cdot 32$ Hz, $z_0 = 1$, we find that the two inflection points are determined by 
$\cos(\omega_2 t) = 0$, {\it i.e.}, $\omega_2 t = \pi/2$ and $3\pi/2$, which yields the above two values. At these inflection points,
the density profiles in momentum space for both spin components will have identical structure (see Fig. \ref{fig-fig5}), which is the same 
as $T/4$ in Fig. \ref{fig-fig2} (b). The inflection points disappear when $2J/\omega_2 < \pi$. This non-sinusoidal dynamics represents the distinctive 
feature of ultracold atoms, because in solid materials the tunneling is determined by the parameters of the barrier, thus can not be modulated in experiments.

Finally, several remarks are in order. (I) This model has some salient advantages as compared with the
previous systems. These condensates are separated from the excitations by a finite gap, which can be
made to be comparable with the BEC temperature, thus can suppress the damping effect induced by the thermal atoms. In this case, the oscillation period 
can be tuned by several order of magnitudes, from millisecond \cite{estimateTmillsecond} to one hundreds of milliseconds, without influencing the other parameters.
This kind of tunability is challenging to be realized using a double-well potential. 
In experiments, these dynamics can be visualized from measurement in spin and momentum spaces, thus is not limited by resolution during 
absorption imaging. (II) The similar physics can be realized by considering various structures. In a cigar-shaped BEC or a condensate with one dimensional SOC, this model will
be reduced to momentum space Josephson effect \cite{hou2018momentum, bersano2018experimental}. In a toroidal potential, this angular momentum Josephson effect
can be used as sensor for rotation \cite{gallemi2016coherent, levy2007ac}. (III) These two levels can also be used to realize the nonlinear Lipkin-Meshkov-Glick model in the form $\mathcal{H} = 
{\bf h} \cdot {\bf \hat{J}} + K \hat{J}_z^2$ \citep{pan1999analytical,ma2011quantum,salvatori2014quantum, luo2017deterministic}, where ${\bf h}= 
{\bf h}(t)$ is the driven Zeeman field and $\hat{J}_x = (a_1^\dagger a_2 + a_2^\dagger a_1)/2$, $\hat{J}_y = (a_1^\dagger a_2 - a_2^\dagger a_1)/2i$ 
and $\hat{J}_z = (a_1^\dagger a_1 - a_2^\dagger a_2)/2$ in Schwinger representation \cite{milburn1997quantum, graefe2007semiclassical}, for squeezed state in quantum 
metrology \cite{pezze2018quantum}. 

To conclude we show that the two degenerate levels in the spin-orbit coupled Bose gases in a trap can be used for 
searching of angular momentum Josephson dynamics. This platform has several advantages as compared with the previous ones, such as suppressed  
damping effect, much wider tunable period and measurement in spin and momentum spaces. We demonstrate this effect using experimental accessible parameters. 
Our results show that this system can be used as a appealing platform for matter wave interferometry and  quantum metrology. Interesting 
applications based on these two levels are promising, such as squeezed condensate and entangled Schr\"odinger cat state, in AMO physics.

\textit{Acknowledgements.} This work is supported by the National Youth Thousand Talents Program (No. KJ2030000001), the USTC start-up funding (No. KY2030000053), the NSFC (No. 11774328) and the  National Key   Research and Development Program of China (No. 2016YFA0301700).

\bibliography{ref}

\begin{thebibliography}{78}%
\makeatletter
\providecommand \@ifxundefined [1]{%
 \@ifx{#1\undefined}
}%
\providecommand \@ifnum [1]{%
 \ifnum #1\expandafter \@firstoftwo
 \else \expandafter \@secondoftwo
 \fi
}%
\providecommand \@ifx [1]{%
 \ifx #1\expandafter \@firstoftwo
 \else \expandafter \@secondoftwo
 \fi
}%
\providecommand \natexlab [1]{#1}%
\providecommand \enquote  [1]{``#1''}%
\providecommand \bibnamefont  [1]{#1}%
\providecommand \bibfnamefont [1]{#1}%
\providecommand \citenamefont [1]{#1}%
\providecommand \href@noop [0]{\@secondoftwo}%
\providecommand \href [0]{\begingroup \@sanitize@url \@href}%
\providecommand \@href[1]{\@@startlink{#1}\@@href}%
\providecommand \@@href[1]{\endgroup#1\@@endlink}%
\providecommand \@sanitize@url [0]{\catcode `\\12\catcode `\$12\catcode
  `\&12\catcode `\#12\catcode `\^12\catcode `\_12\catcode `\%12\relax}%
\providecommand \@@startlink[1]{}%
\providecommand \@@endlink[0]{}%
\providecommand \url  [0]{\begingroup\@sanitize@url \@url }%
\providecommand \@url [1]{\endgroup\@href {#1}{\urlprefix }}%
\providecommand \urlprefix  [0]{URL }%
\providecommand \Eprint [0]{\href }%
\providecommand \doibase [0]{http://dx.doi.org/}%
\providecommand \selectlanguage [0]{\@gobble}%
\providecommand \bibinfo  [0]{\@secondoftwo}%
\providecommand \bibfield  [0]{\@secondoftwo}%
\providecommand \translation [1]{[#1]}%
\providecommand \BibitemOpen [0]{}%
\providecommand \bibitemStop [0]{}%
\providecommand \bibitemNoStop [0]{.\EOS\space}%
\providecommand \EOS [0]{\spacefactor3000\relax}%
\providecommand \BibitemShut  [1]{\csname bibitem#1\endcsname}%
\let\auto@bib@innerbib\@empty
\bibitem [{\citenamefont {Josephson}(1962)}]{josephson1962possible}%
  \BibitemOpen
  \bibfield  {author} {\bibinfo {author} {\bibfnamefont {B.~D.}\ \bibnamefont
  {Josephson}},\ }\href@noop {} {\bibfield  {journal} {\bibinfo  {journal}
  {Phys. Rev. Lett.}\ }\textbf {\bibinfo {volume} {1}},\ \bibinfo {pages} {251}
  (\bibinfo {year} {1962})}\BibitemShut {NoStop}%
\bibitem [{\citenamefont {Josephson}(1974)}]{josephson1974discovery}%
  \BibitemOpen
  \bibfield  {author} {\bibinfo {author} {\bibfnamefont {B.~D.}\ \bibnamefont
  {Josephson}},\ }\href@noop {} {\bibfield  {journal} {\bibinfo  {journal}
  {Europhys. News}\ }\textbf {\bibinfo {volume} {5}},\ \bibinfo {pages} {1}
  (\bibinfo {year} {1974})}\BibitemShut {NoStop}%
\bibitem [{\citenamefont {Feynman}\ \emph {et~al.}(1965)\citenamefont
  {Feynman}, \citenamefont {Leighton},\ and\ \citenamefont
  {Sands}}]{feynman1965lectures}%
  \BibitemOpen
  \bibfield  {author} {\bibinfo {author} {\bibfnamefont {R.~P.}\ \bibnamefont
  {Feynman}}, \bibinfo {author} {\bibfnamefont {R.~B.}\ \bibnamefont
  {Leighton}}, \ and\ \bibinfo {author} {\bibfnamefont {M.}~\bibnamefont
  {Sands}},\ }\href@noop {} {\enquote {\bibinfo {title} {Lectures on physics,
  vol. iii},}\ } (\bibinfo {year} {1965})\BibitemShut {NoStop}%
\bibitem [{\citenamefont {Bloch}(1970)}]{bloch1970josephson}%
  \BibitemOpen
  \bibfield  {author} {\bibinfo {author} {\bibfnamefont {F.}~\bibnamefont
  {Bloch}},\ }\href@noop {} {\bibfield  {journal} {\bibinfo  {journal} {Phys.
  Rev. B}\ }\textbf {\bibinfo {volume} {2}},\ \bibinfo {pages} {109} (\bibinfo
  {year} {1970})}\BibitemShut {NoStop}%
\bibitem [{\citenamefont {Anderson}\ and\ \citenamefont
  {Rowell}(1963)}]{anderson1963probable}%
  \BibitemOpen
  \bibfield  {author} {\bibinfo {author} {\bibfnamefont {P.~W.}\ \bibnamefont
  {Anderson}}\ and\ \bibinfo {author} {\bibfnamefont {J.~M.}\ \bibnamefont
  {Rowell}},\ }\href@noop {} {\bibfield  {journal} {\bibinfo  {journal} {Phys.
  Rev. Lett.}\ }\textbf {\bibinfo {volume} {10}},\ \bibinfo {pages} {230}
  (\bibinfo {year} {1963})}\BibitemShut {NoStop}%
\bibitem [{\citenamefont {B{\"u}ttiker}\ \emph {et~al.}(1983)\citenamefont
  {B{\"u}ttiker}, \citenamefont {Imry},\ and\ \citenamefont
  {Landauer}}]{buttiker1983josephson}%
  \BibitemOpen
  \bibfield  {author} {\bibinfo {author} {\bibfnamefont {M.}~\bibnamefont
  {B{\"u}ttiker}}, \bibinfo {author} {\bibfnamefont {Y.}~\bibnamefont {Imry}},
  \ and\ \bibinfo {author} {\bibfnamefont {R.}~\bibnamefont {Landauer}},\
  }\href@noop {} {\bibfield  {journal} {\bibinfo  {journal} {Phys. Lett. A}\
  }\textbf {\bibinfo {volume} {96}},\ \bibinfo {pages} {365} (\bibinfo {year}
  {1983})}\BibitemShut {NoStop}%
\bibitem [{\citenamefont {Sukhatme}\ \emph {et~al.}(2001)\citenamefont
  {Sukhatme}, \citenamefont {Mukharsky}, \citenamefont {Chui},\ and\
  \citenamefont {Pearson}}]{sukhatme2001observation}%
  \BibitemOpen
  \bibfield  {author} {\bibinfo {author} {\bibfnamefont {K.}~\bibnamefont
  {Sukhatme}}, \bibinfo {author} {\bibfnamefont {Y.}~\bibnamefont {Mukharsky}},
  \bibinfo {author} {\bibfnamefont {T.}~\bibnamefont {Chui}}, \ and\ \bibinfo
  {author} {\bibfnamefont {D.}~\bibnamefont {Pearson}},\ }\href@noop {}
  {\bibfield  {journal} {\bibinfo  {journal} {Nature}\ }\textbf {\bibinfo
  {volume} {411}},\ \bibinfo {pages} {280} (\bibinfo {year}
  {2001})}\BibitemShut {NoStop}%
\bibitem [{\citenamefont {Avenel}\ and\ \citenamefont
  {Varoquaux}(1988)}]{avenel1988josephson}%
  \BibitemOpen
  \bibfield  {author} {\bibinfo {author} {\bibfnamefont {O.}~\bibnamefont
  {Avenel}}\ and\ \bibinfo {author} {\bibfnamefont {E.}~\bibnamefont
  {Varoquaux}},\ }\href@noop {} {\bibfield  {journal} {\bibinfo  {journal}
  {Phys. Rev. Lett.}\ }\textbf {\bibinfo {volume} {60}},\ \bibinfo {pages}
  {416} (\bibinfo {year} {1988})}\BibitemShut {NoStop}%
\bibitem [{\citenamefont {Albiez}\ \emph {et~al.}(2005)\citenamefont {Albiez},
  \citenamefont {Gati}, \citenamefont {F{\"o}lling}, \citenamefont {Hunsmann},
  \citenamefont {Cristiani},\ and\ \citenamefont
  {Oberthaler}}]{albiez2005direct}%
  \BibitemOpen
  \bibfield  {author} {\bibinfo {author} {\bibfnamefont {M.}~\bibnamefont
  {Albiez}}, \bibinfo {author} {\bibfnamefont {R.}~\bibnamefont {Gati}},
  \bibinfo {author} {\bibfnamefont {J.}~\bibnamefont {F{\"o}lling}}, \bibinfo
  {author} {\bibfnamefont {S.}~\bibnamefont {Hunsmann}}, \bibinfo {author}
  {\bibfnamefont {M.}~\bibnamefont {Cristiani}}, \ and\ \bibinfo {author}
  {\bibfnamefont {M.~K.}\ \bibnamefont {Oberthaler}},\ }\href@noop {}
  {\bibfield  {journal} {\bibinfo  {journal} {Phys. Rev. Lett.}\ }\textbf
  {\bibinfo {volume} {95}},\ \bibinfo {pages} {010402} (\bibinfo {year}
  {2005})}\BibitemShut {NoStop}%
\bibitem [{\citenamefont {Levy}\ \emph {et~al.}(2007)\citenamefont {Levy},
  \citenamefont {Lahoud}, \citenamefont {Shomroni},\ and\ \citenamefont
  {Steinhauer}}]{levy2007ac}%
  \BibitemOpen
  \bibfield  {author} {\bibinfo {author} {\bibfnamefont {S.}~\bibnamefont
  {Levy}}, \bibinfo {author} {\bibfnamefont {E.}~\bibnamefont {Lahoud}},
  \bibinfo {author} {\bibfnamefont {I.}~\bibnamefont {Shomroni}}, \ and\
  \bibinfo {author} {\bibfnamefont {J.}~\bibnamefont {Steinhauer}},\
  }\href@noop {} {\bibfield  {journal} {\bibinfo  {journal} {Nature}\ }\textbf
  {\bibinfo {volume} {449}},\ \bibinfo {pages} {579} (\bibinfo {year}
  {2007})}\BibitemShut {NoStop}%
\bibitem [{\citenamefont {Cataliotti}\ \emph {et~al.}(2001)\citenamefont
  {Cataliotti}, \citenamefont {Burger}, \citenamefont {Fort}, \citenamefont
  {Maddaloni}, \citenamefont {Minardi}, \citenamefont {Trombettoni},
  \citenamefont {Smerzi},\ and\ \citenamefont
  {Inguscio}}]{cataliotti2001josephson}%
  \BibitemOpen
  \bibfield  {author} {\bibinfo {author} {\bibfnamefont {F.}~\bibnamefont
  {Cataliotti}}, \bibinfo {author} {\bibfnamefont {S.}~\bibnamefont {Burger}},
  \bibinfo {author} {\bibfnamefont {C.}~\bibnamefont {Fort}}, \bibinfo {author}
  {\bibfnamefont {P.}~\bibnamefont {Maddaloni}}, \bibinfo {author}
  {\bibfnamefont {F.}~\bibnamefont {Minardi}}, \bibinfo {author} {\bibfnamefont
  {A.}~\bibnamefont {Trombettoni}}, \bibinfo {author} {\bibfnamefont
  {A.}~\bibnamefont {Smerzi}}, \ and\ \bibinfo {author} {\bibfnamefont
  {M.}~\bibnamefont {Inguscio}},\ }\href@noop {} {\bibfield  {journal}
  {\bibinfo  {journal} {Science}\ }\textbf {\bibinfo {volume} {293}},\ \bibinfo
  {pages} {843} (\bibinfo {year} {2001})}\BibitemShut {NoStop}%
\bibitem [{\citenamefont {Lagoudakis}\ \emph {et~al.}(2010)\citenamefont
  {Lagoudakis}, \citenamefont {Pietka}, \citenamefont {Wouters}, \citenamefont
  {Andr{\'e}},\ and\ \citenamefont
  {Deveaud-Pl{\'e}dran}}]{lagoudakis2010coherent}%
  \BibitemOpen
  \bibfield  {author} {\bibinfo {author} {\bibfnamefont {K.}~\bibnamefont
  {Lagoudakis}}, \bibinfo {author} {\bibfnamefont {B.}~\bibnamefont {Pietka}},
  \bibinfo {author} {\bibfnamefont {M.}~\bibnamefont {Wouters}}, \bibinfo
  {author} {\bibfnamefont {R.}~\bibnamefont {Andr{\'e}}}, \ and\ \bibinfo
  {author} {\bibfnamefont {B.}~\bibnamefont {Deveaud-Pl{\'e}dran}},\
  }\href@noop {} {\bibfield  {journal} {\bibinfo  {journal} {Phys. Rev. Lett.}\
  }\textbf {\bibinfo {volume} {105}},\ \bibinfo {pages} {120403} (\bibinfo
  {year} {2010})}\BibitemShut {NoStop}%
\bibitem [{\citenamefont {Abbarchi}\ \emph {et~al.}(2013)\citenamefont
  {Abbarchi}, \citenamefont {Amo}, \citenamefont {Sala}, \citenamefont
  {Solnyshkov}, \citenamefont {Flayac}, \citenamefont {Ferrier}, \citenamefont
  {Sagnes}, \citenamefont {Galopin}, \citenamefont {Lema{\^\i}tre},
  \citenamefont {Malpuech} \emph {et~al.}}]{abbarchi2013macroscopic}%
  \BibitemOpen
  \bibfield  {author} {\bibinfo {author} {\bibfnamefont {M.}~\bibnamefont
  {Abbarchi}}, \bibinfo {author} {\bibfnamefont {A.}~\bibnamefont {Amo}},
  \bibinfo {author} {\bibfnamefont {V.}~\bibnamefont {Sala}}, \bibinfo {author}
  {\bibfnamefont {D.}~\bibnamefont {Solnyshkov}}, \bibinfo {author}
  {\bibfnamefont {H.}~\bibnamefont {Flayac}}, \bibinfo {author} {\bibfnamefont
  {L.}~\bibnamefont {Ferrier}}, \bibinfo {author} {\bibfnamefont
  {I.}~\bibnamefont {Sagnes}}, \bibinfo {author} {\bibfnamefont
  {E.}~\bibnamefont {Galopin}}, \bibinfo {author} {\bibfnamefont
  {A.}~\bibnamefont {Lema{\^\i}tre}}, \bibinfo {author} {\bibfnamefont
  {G.}~\bibnamefont {Malpuech}},  \emph {et~al.},\ }\href@noop {} {\bibfield
  {journal} {\bibinfo  {journal} {Nat. Phys.}\ }\textbf {\bibinfo {volume}
  {9}},\ \bibinfo {pages} {275} (\bibinfo {year} {2013})}\BibitemShut {NoStop}%
\bibitem [{\citenamefont {Valtolina}\ \emph {et~al.}(2015)\citenamefont
  {Valtolina}, \citenamefont {Burchianti}, \citenamefont {Amico}, \citenamefont
  {Neri}, \citenamefont {Xhani}, \citenamefont {Seman}, \citenamefont
  {Trombettoni}, \citenamefont {Smerzi}, \citenamefont {Zaccanti},
  \citenamefont {Inguscio} \emph {et~al.}}]{valtolina2015josephson}%
  \BibitemOpen
  \bibfield  {author} {\bibinfo {author} {\bibfnamefont {G.}~\bibnamefont
  {Valtolina}}, \bibinfo {author} {\bibfnamefont {A.}~\bibnamefont
  {Burchianti}}, \bibinfo {author} {\bibfnamefont {A.}~\bibnamefont {Amico}},
  \bibinfo {author} {\bibfnamefont {E.}~\bibnamefont {Neri}}, \bibinfo {author}
  {\bibfnamefont {K.}~\bibnamefont {Xhani}}, \bibinfo {author} {\bibfnamefont
  {J.~A.}\ \bibnamefont {Seman}}, \bibinfo {author} {\bibfnamefont
  {A.}~\bibnamefont {Trombettoni}}, \bibinfo {author} {\bibfnamefont
  {A.}~\bibnamefont {Smerzi}}, \bibinfo {author} {\bibfnamefont
  {M.}~\bibnamefont {Zaccanti}}, \bibinfo {author} {\bibfnamefont
  {M.}~\bibnamefont {Inguscio}},  \emph {et~al.},\ }\href@noop {} {\bibfield
  {journal} {\bibinfo  {journal} {Science}\ }\textbf {\bibinfo {volume}
  {350}},\ \bibinfo {pages} {1505} (\bibinfo {year} {2015})}\BibitemShut
  {NoStop}%
\bibitem [{\citenamefont {Zimmerman}\ and\ \citenamefont
  {Silver}(1966)}]{zimmerman1966macroscopic}%
  \BibitemOpen
  \bibfield  {author} {\bibinfo {author} {\bibfnamefont {J.}~\bibnamefont
  {Zimmerman}}\ and\ \bibinfo {author} {\bibfnamefont {A.}~\bibnamefont
  {Silver}},\ }\href@noop {} {\bibfield  {journal} {\bibinfo  {journal} {Phys.
  Rev.}\ }\textbf {\bibinfo {volume} {141}},\ \bibinfo {pages} {367} (\bibinfo
  {year} {1966})}\BibitemShut {NoStop}%
\bibitem [{\citenamefont {Koch}\ \emph {et~al.}(1987)\citenamefont {Koch},
  \citenamefont {Umbach}, \citenamefont {Clark}, \citenamefont {Chaudhari},\
  and\ \citenamefont {Laibowitz}}]{koch1987quantum}%
  \BibitemOpen
  \bibfield  {author} {\bibinfo {author} {\bibfnamefont {R.}~\bibnamefont
  {Koch}}, \bibinfo {author} {\bibfnamefont {C.}~\bibnamefont {Umbach}},
  \bibinfo {author} {\bibfnamefont {G.}~\bibnamefont {Clark}}, \bibinfo
  {author} {\bibfnamefont {P.}~\bibnamefont {Chaudhari}}, \ and\ \bibinfo
  {author} {\bibfnamefont {R.}~\bibnamefont {Laibowitz}},\ }\href@noop {}
  {\bibfield  {journal} {\bibinfo  {journal} {Appl. Phys. Lett.}\ }\textbf
  {\bibinfo {volume} {51}},\ \bibinfo {pages} {200} (\bibinfo {year}
  {1987})}\BibitemShut {NoStop}%
\bibitem [{\citenamefont {Fagaly}(2006)}]{fagaly2006superconducting}%
  \BibitemOpen
  \bibfield  {author} {\bibinfo {author} {\bibfnamefont {R.}~\bibnamefont
  {Fagaly}},\ }\href@noop {} {\bibfield  {journal} {\bibinfo  {journal} {Rev.
  Sci. Instrum.}\ }\textbf {\bibinfo {volume} {77}},\ \bibinfo {pages} {101101}
  (\bibinfo {year} {2006})}\BibitemShut {NoStop}%
\bibitem [{\citenamefont {Ryu}\ \emph {et~al.}(2013)\citenamefont {Ryu},
  \citenamefont {Blackburn}, \citenamefont {Blinova},\ and\ \citenamefont
  {Boshier}}]{ryu2013experimental}%
  \BibitemOpen
  \bibfield  {author} {\bibinfo {author} {\bibfnamefont {C.}~\bibnamefont
  {Ryu}}, \bibinfo {author} {\bibfnamefont {P.}~\bibnamefont {Blackburn}},
  \bibinfo {author} {\bibfnamefont {A.}~\bibnamefont {Blinova}}, \ and\
  \bibinfo {author} {\bibfnamefont {M.}~\bibnamefont {Boshier}},\ }\href@noop
  {} {\bibfield  {journal} {\bibinfo  {journal} {Phys. Rev. Lett.}\ }\textbf
  {\bibinfo {volume} {111}},\ \bibinfo {pages} {205301} (\bibinfo {year}
  {2013})}\BibitemShut {NoStop}%
\bibitem [{\citenamefont {Gallop}(2003)}]{gallop2003squids}%
  \BibitemOpen
  \bibfield  {author} {\bibinfo {author} {\bibfnamefont {J.}~\bibnamefont
  {Gallop}},\ }\href@noop {} {\bibfield  {journal} {\bibinfo  {journal}
  {Supercond. Sci. Technol.}\ }\textbf {\bibinfo {volume} {16}},\ \bibinfo
  {pages} {1575} (\bibinfo {year} {2003})}\BibitemShut {NoStop}%
\bibitem [{\citenamefont {Ofek}\ \emph {et~al.}(2016)\citenamefont {Ofek},
  \citenamefont {Petrenko}, \citenamefont {Heeres}, \citenamefont {Reinhold},
  \citenamefont {Leghtas}, \citenamefont {Vlastakis}, \citenamefont {Liu},
  \citenamefont {Frunzio}, \citenamefont {Girvin}, \citenamefont {Jiang} \emph
  {et~al.}}]{ofek2016extending}%
  \BibitemOpen
  \bibfield  {author} {\bibinfo {author} {\bibfnamefont {N.}~\bibnamefont
  {Ofek}}, \bibinfo {author} {\bibfnamefont {A.}~\bibnamefont {Petrenko}},
  \bibinfo {author} {\bibfnamefont {R.}~\bibnamefont {Heeres}}, \bibinfo
  {author} {\bibfnamefont {P.}~\bibnamefont {Reinhold}}, \bibinfo {author}
  {\bibfnamefont {Z.}~\bibnamefont {Leghtas}}, \bibinfo {author} {\bibfnamefont
  {B.}~\bibnamefont {Vlastakis}}, \bibinfo {author} {\bibfnamefont
  {Y.}~\bibnamefont {Liu}}, \bibinfo {author} {\bibfnamefont {L.}~\bibnamefont
  {Frunzio}}, \bibinfo {author} {\bibfnamefont {S.}~\bibnamefont {Girvin}},
  \bibinfo {author} {\bibfnamefont {L.}~\bibnamefont {Jiang}},  \emph
  {et~al.},\ }\href@noop {} {\bibfield  {journal} {\bibinfo  {journal}
  {Nature}\ }\textbf {\bibinfo {volume} {536}},\ \bibinfo {pages} {441}
  (\bibinfo {year} {2016})}\BibitemShut {NoStop}%
\bibitem [{\citenamefont {C{\'o}rcoles}\ \emph {et~al.}(2015)\citenamefont
  {C{\'o}rcoles}, \citenamefont {Magesan}, \citenamefont {Srinivasan},
  \citenamefont {Cross}, \citenamefont {Steffen}, \citenamefont {Gambetta},\
  and\ \citenamefont {Chow}}]{corcoles2015demonstration}%
  \BibitemOpen
  \bibfield  {author} {\bibinfo {author} {\bibfnamefont {A.~D.}\ \bibnamefont
  {C{\'o}rcoles}}, \bibinfo {author} {\bibfnamefont {E.}~\bibnamefont
  {Magesan}}, \bibinfo {author} {\bibfnamefont {S.~J.}\ \bibnamefont
  {Srinivasan}}, \bibinfo {author} {\bibfnamefont {A.~W.}\ \bibnamefont
  {Cross}}, \bibinfo {author} {\bibfnamefont {M.}~\bibnamefont {Steffen}},
  \bibinfo {author} {\bibfnamefont {J.~M.}\ \bibnamefont {Gambetta}}, \ and\
  \bibinfo {author} {\bibfnamefont {J.~M.}\ \bibnamefont {Chow}},\ }\href@noop
  {} {\bibfield  {journal} {\bibinfo  {journal} {Nat. Commun.}\ }\textbf
  {\bibinfo {volume} {6}},\ \bibinfo {pages} {6979} (\bibinfo {year}
  {2015})}\BibitemShut {NoStop}%
\bibitem [{\citenamefont {Kelly}\ \emph {et~al.}(2015)\citenamefont {Kelly},
  \citenamefont {Barends}, \citenamefont {Fowler}, \citenamefont {Megrant},
  \citenamefont {Jeffrey}, \citenamefont {White}, \citenamefont {Sank},
  \citenamefont {Mutus}, \citenamefont {Campbell}, \citenamefont {Chen} \emph
  {et~al.}}]{kelly2015state}%
  \BibitemOpen
  \bibfield  {author} {\bibinfo {author} {\bibfnamefont {J.}~\bibnamefont
  {Kelly}}, \bibinfo {author} {\bibfnamefont {R.}~\bibnamefont {Barends}},
  \bibinfo {author} {\bibfnamefont {A.~G.}\ \bibnamefont {Fowler}}, \bibinfo
  {author} {\bibfnamefont {A.}~\bibnamefont {Megrant}}, \bibinfo {author}
  {\bibfnamefont {E.}~\bibnamefont {Jeffrey}}, \bibinfo {author} {\bibfnamefont
  {T.~C.}\ \bibnamefont {White}}, \bibinfo {author} {\bibfnamefont
  {D.}~\bibnamefont {Sank}}, \bibinfo {author} {\bibfnamefont {J.~Y.}\
  \bibnamefont {Mutus}}, \bibinfo {author} {\bibfnamefont {B.}~\bibnamefont
  {Campbell}}, \bibinfo {author} {\bibfnamefont {Y.}~\bibnamefont {Chen}},
  \emph {et~al.},\ }\href@noop {} {\bibfield  {journal} {\bibinfo  {journal}
  {Nature}\ }\textbf {\bibinfo {volume} {519}},\ \bibinfo {pages} {66}
  (\bibinfo {year} {2015})}\BibitemShut {NoStop}%
\bibitem [{\citenamefont {Ozyuzer}\ \emph {et~al.}(2007)\citenamefont
  {Ozyuzer}, \citenamefont {Koshelev}, \citenamefont {Kurter}, \citenamefont
  {Gopalsami}, \citenamefont {Li}, \citenamefont {Tachiki}, \citenamefont
  {Kadowaki}, \citenamefont {Yamamoto}, \citenamefont {Minami}, \citenamefont
  {Yamaguchi} \emph {et~al.}}]{ozyuzer2007emission}%
  \BibitemOpen
  \bibfield  {author} {\bibinfo {author} {\bibfnamefont {L.}~\bibnamefont
  {Ozyuzer}}, \bibinfo {author} {\bibfnamefont {A.~E.}\ \bibnamefont
  {Koshelev}}, \bibinfo {author} {\bibfnamefont {C.}~\bibnamefont {Kurter}},
  \bibinfo {author} {\bibfnamefont {N.}~\bibnamefont {Gopalsami}}, \bibinfo
  {author} {\bibfnamefont {Q.}~\bibnamefont {Li}}, \bibinfo {author}
  {\bibfnamefont {M.}~\bibnamefont {Tachiki}}, \bibinfo {author} {\bibfnamefont
  {K.}~\bibnamefont {Kadowaki}}, \bibinfo {author} {\bibfnamefont
  {T.}~\bibnamefont {Yamamoto}}, \bibinfo {author} {\bibfnamefont
  {H.}~\bibnamefont {Minami}}, \bibinfo {author} {\bibfnamefont
  {H.}~\bibnamefont {Yamaguchi}},  \emph {et~al.},\ }\href@noop {} {\bibfield
  {journal} {\bibinfo  {journal} {Science}\ }\textbf {\bibinfo {volume}
  {318}},\ \bibinfo {pages} {1291} (\bibinfo {year} {2007})}\BibitemShut
  {NoStop}%
\bibitem [{\citenamefont {Tachiki}\ \emph {et~al.}(2005)\citenamefont
  {Tachiki}, \citenamefont {Iizuka}, \citenamefont {Minami}, \citenamefont
  {Tejima},\ and\ \citenamefont {Nakamura}}]{tachiki2005emission}%
  \BibitemOpen
  \bibfield  {author} {\bibinfo {author} {\bibfnamefont {M.}~\bibnamefont
  {Tachiki}}, \bibinfo {author} {\bibfnamefont {M.}~\bibnamefont {Iizuka}},
  \bibinfo {author} {\bibfnamefont {K.}~\bibnamefont {Minami}}, \bibinfo
  {author} {\bibfnamefont {S.}~\bibnamefont {Tejima}}, \ and\ \bibinfo {author}
  {\bibfnamefont {H.}~\bibnamefont {Nakamura}},\ }\href@noop {} {\bibfield
  {journal} {\bibinfo  {journal} {Phys. Rev. B}\ }\textbf {\bibinfo {volume}
  {71}},\ \bibinfo {pages} {134515} (\bibinfo {year} {2005})}\BibitemShut
  {NoStop}%
\bibitem [{\citenamefont {Shapiro}(1963)}]{shapiro1963josephson}%
  \BibitemOpen
  \bibfield  {author} {\bibinfo {author} {\bibfnamefont {S.}~\bibnamefont
  {Shapiro}},\ }\href@noop {} {\bibfield  {journal} {\bibinfo  {journal} {Phys.
  Rev. Lett.}\ }\textbf {\bibinfo {volume} {11}},\ \bibinfo {pages} {80}
  (\bibinfo {year} {1963})}\BibitemShut {NoStop}%
\bibitem [{\citenamefont {Kleiner}\ \emph {et~al.}(1992)\citenamefont
  {Kleiner}, \citenamefont {Steinmeyer}, \citenamefont {Kunkel},\ and\
  \citenamefont {M{\"u}ller}}]{kleiner1992intrinsic}%
  \BibitemOpen
  \bibfield  {author} {\bibinfo {author} {\bibfnamefont {R.}~\bibnamefont
  {Kleiner}}, \bibinfo {author} {\bibfnamefont {F.}~\bibnamefont {Steinmeyer}},
  \bibinfo {author} {\bibfnamefont {G.}~\bibnamefont {Kunkel}}, \ and\ \bibinfo
  {author} {\bibfnamefont {P.}~\bibnamefont {M{\"u}ller}},\ }\href@noop {}
  {\bibfield  {journal} {\bibinfo  {journal} {Phys. Rev. Lett.}\ }\textbf
  {\bibinfo {volume} {68}},\ \bibinfo {pages} {2394} (\bibinfo {year}
  {1992})}\BibitemShut {NoStop}%
\bibitem [{\citenamefont {Smerzi}\ \emph {et~al.}(1997)\citenamefont {Smerzi},
  \citenamefont {Fantoni}, \citenamefont {Giovanazzi},\ and\ \citenamefont
  {Shenoy}}]{smerzi1997quantum}%
  \BibitemOpen
  \bibfield  {author} {\bibinfo {author} {\bibfnamefont {A.}~\bibnamefont
  {Smerzi}}, \bibinfo {author} {\bibfnamefont {S.}~\bibnamefont {Fantoni}},
  \bibinfo {author} {\bibfnamefont {S.}~\bibnamefont {Giovanazzi}}, \ and\
  \bibinfo {author} {\bibfnamefont {S.}~\bibnamefont {Shenoy}},\ }\href@noop {}
  {\bibfield  {journal} {\bibinfo  {journal} {Phys. Rev. Lett.}\ }\textbf
  {\bibinfo {volume} {79}},\ \bibinfo {pages} {4950} (\bibinfo {year}
  {1997})}\BibitemShut {NoStop}%
\bibitem [{\citenamefont {Javanainen}(1986)}]{javanainen1986oscillatory}%
  \BibitemOpen
  \bibfield  {author} {\bibinfo {author} {\bibfnamefont {J.}~\bibnamefont
  {Javanainen}},\ }\href@noop {} {\bibfield  {journal} {\bibinfo  {journal}
  {Phys. Rev. Lett.}\ }\textbf {\bibinfo {volume} {57}},\ \bibinfo {pages}
  {3164} (\bibinfo {year} {1986})}\BibitemShut {NoStop}%
\bibitem [{\citenamefont {Shin}\ \emph {et~al.}(2005)\citenamefont {Shin},
  \citenamefont {Jo}, \citenamefont {Saba}, \citenamefont {Pasquini},
  \citenamefont {Ketterle},\ and\ \citenamefont {Pritchard}}]{shin2005optical}%
  \BibitemOpen
  \bibfield  {author} {\bibinfo {author} {\bibfnamefont {Y.}~\bibnamefont
  {Shin}}, \bibinfo {author} {\bibfnamefont {G.-B.}\ \bibnamefont {Jo}},
  \bibinfo {author} {\bibfnamefont {M.}~\bibnamefont {Saba}}, \bibinfo {author}
  {\bibfnamefont {T.}~\bibnamefont {Pasquini}}, \bibinfo {author}
  {\bibfnamefont {W.}~\bibnamefont {Ketterle}}, \ and\ \bibinfo {author}
  {\bibfnamefont {D.~E.}\ \bibnamefont {Pritchard}},\ }\href@noop {} {\bibfield
   {journal} {\bibinfo  {journal} {Phys. Rev. Lett.}\ }\textbf {\bibinfo
  {volume} {95}},\ \bibinfo {pages} {170402} (\bibinfo {year}
  {2005})}\BibitemShut {NoStop}%
\bibitem [{\citenamefont {Zibold}\ \emph {et~al.}(2010)\citenamefont {Zibold},
  \citenamefont {Nicklas}, \citenamefont {Gross},\ and\ \citenamefont
  {Oberthaler}}]{zibold2010classical}%
  \BibitemOpen
  \bibfield  {author} {\bibinfo {author} {\bibfnamefont {T.}~\bibnamefont
  {Zibold}}, \bibinfo {author} {\bibfnamefont {E.}~\bibnamefont {Nicklas}},
  \bibinfo {author} {\bibfnamefont {C.}~\bibnamefont {Gross}}, \ and\ \bibinfo
  {author} {\bibfnamefont {M.~K.}\ \bibnamefont {Oberthaler}},\ }\href@noop {}
  {\bibfield  {journal} {\bibinfo  {journal} {Phys. Rev. Lett.}\ }\textbf
  {\bibinfo {volume} {105}},\ \bibinfo {pages} {204101} (\bibinfo {year}
  {2010})}\BibitemShut {NoStop}%
\bibitem [{\citenamefont {Chang}\ \emph {et~al.}(2005)\citenamefont {Chang},
  \citenamefont {Qin}, \citenamefont {Zhang}, \citenamefont {You},\ and\
  \citenamefont {Chapman}}]{chang2005coherent}%
  \BibitemOpen
  \bibfield  {author} {\bibinfo {author} {\bibfnamefont {M.-S.}\ \bibnamefont
  {Chang}}, \bibinfo {author} {\bibfnamefont {Q.}~\bibnamefont {Qin}}, \bibinfo
  {author} {\bibfnamefont {W.}~\bibnamefont {Zhang}}, \bibinfo {author}
  {\bibfnamefont {L.}~\bibnamefont {You}}, \ and\ \bibinfo {author}
  {\bibfnamefont {M.~S.}\ \bibnamefont {Chapman}},\ }\href@noop {} {\bibfield
  {journal} {\bibinfo  {journal} {Nat. Phys.}\ }\textbf {\bibinfo {volume}
  {1}},\ \bibinfo {pages} {111} (\bibinfo {year} {2005})}\BibitemShut {NoStop}%
\bibitem [{\citenamefont {Hou}\ \emph {et~al.}(2018)\citenamefont {Hou},
  \citenamefont {Luo}, \citenamefont {Sun}, \citenamefont {Bersano},
  \citenamefont {Gokhroo}, \citenamefont {Mossman}, \citenamefont {Engels},\
  and\ \citenamefont {Zhang}}]{hou2018momentum}%
  \BibitemOpen
  \bibfield  {author} {\bibinfo {author} {\bibfnamefont {J.}~\bibnamefont
  {Hou}}, \bibinfo {author} {\bibfnamefont {X.-W.}\ \bibnamefont {Luo}},
  \bibinfo {author} {\bibfnamefont {K.}~\bibnamefont {Sun}}, \bibinfo {author}
  {\bibfnamefont {T.}~\bibnamefont {Bersano}}, \bibinfo {author} {\bibfnamefont
  {V.}~\bibnamefont {Gokhroo}}, \bibinfo {author} {\bibfnamefont
  {S.}~\bibnamefont {Mossman}}, \bibinfo {author} {\bibfnamefont
  {P.}~\bibnamefont {Engels}}, \ and\ \bibinfo {author} {\bibfnamefont
  {C.}~\bibnamefont {Zhang}},\ }\href@noop {} {\bibfield  {journal} {\bibinfo
  {journal} {Phys.Rev.Lett}\ }\textbf {\bibinfo {volume} {120}},\ \bibinfo
  {pages} {120401} (\bibinfo {year} {2018})}\BibitemShut {NoStop}%
\bibitem [{\citenamefont {Bersano}\ \emph {et~al.}(2018)\citenamefont
  {Bersano}, \citenamefont {Hou}, \citenamefont {Mossman}, \citenamefont
  {Gokhroo}, \citenamefont {Luo}, \citenamefont {Sun}, \citenamefont {Zhang},\
  and\ \citenamefont {Engels}}]{bersano2018experimental}%
  \BibitemOpen
  \bibfield  {author} {\bibinfo {author} {\bibfnamefont {T.~M.}\ \bibnamefont
  {Bersano}}, \bibinfo {author} {\bibfnamefont {J.}~\bibnamefont {Hou}},
  \bibinfo {author} {\bibfnamefont {S.}~\bibnamefont {Mossman}}, \bibinfo
  {author} {\bibfnamefont {V.}~\bibnamefont {Gokhroo}}, \bibinfo {author}
  {\bibfnamefont {X.-W.}\ \bibnamefont {Luo}}, \bibinfo {author} {\bibfnamefont
  {K.}~\bibnamefont {Sun}}, \bibinfo {author} {\bibfnamefont {C.}~\bibnamefont
  {Zhang}}, \ and\ \bibinfo {author} {\bibfnamefont {P.}~\bibnamefont
  {Engels}},\ }\href@noop {} {\bibfield  {journal} {\bibinfo  {journal} {arXiv
  preprint arXiv:1809.01208}\ } (\bibinfo {year} {2018})}\BibitemShut {NoStop}%
\bibitem [{\citenamefont {Gallemí}\ \emph {et~al.}()\citenamefont {Gallemí},
  \citenamefont {Mateo}, \citenamefont {Mayol},\ and\ \citenamefont
  {Guilleumas}}]{gallemi2016coherent}%
  \BibitemOpen
  \bibfield  {author} {\bibinfo {author} {\bibfnamefont {A.}~\bibnamefont
  {Gallemí}}, \bibinfo {author} {\bibfnamefont {A.~M.}\ \bibnamefont {Mateo}},
  \bibinfo {author} {\bibfnamefont {R.}~\bibnamefont {Mayol}}, \ and\ \bibinfo
  {author} {\bibfnamefont {M.}~\bibnamefont {Guilleumas}},\ }\href {\doibase
  doi:10.1088/1367-2630/18/1/015003} {\bibfield  {journal} {\bibinfo  {journal}
  {New. J. Phys}\ }\textbf {\bibinfo {volume} {18}},\ \bibinfo {pages}
  {015003}}\BibitemShut {NoStop}%
\bibitem [{\citenamefont {Zapata}\ \emph {et~al.}(1998)\citenamefont {Zapata},
  \citenamefont {Sols},\ and\ \citenamefont {Leggett}}]{zapata1998josephson}%
  \BibitemOpen
  \bibfield  {author} {\bibinfo {author} {\bibfnamefont {I.}~\bibnamefont
  {Zapata}}, \bibinfo {author} {\bibfnamefont {F.}~\bibnamefont {Sols}}, \ and\
  \bibinfo {author} {\bibfnamefont {A.~J.}\ \bibnamefont {Leggett}},\
  }\href@noop {} {\bibfield  {journal} {\bibinfo  {journal} {Phys. Rev. A}\
  }\textbf {\bibinfo {volume} {57}},\ \bibinfo {pages} {R28} (\bibinfo {year}
  {1998})}\BibitemShut {NoStop}%
\bibitem [{\citenamefont {Polo}\ \emph {et~al.}(2018)\citenamefont {Polo},
  \citenamefont {Ahufinger}, \citenamefont {Hekking},\ and\ \citenamefont
  {Minguzzi}}]{polo2018damping}%
  \BibitemOpen
  \bibfield  {author} {\bibinfo {author} {\bibfnamefont {J.}~\bibnamefont
  {Polo}}, \bibinfo {author} {\bibfnamefont {V.}~\bibnamefont {Ahufinger}},
  \bibinfo {author} {\bibfnamefont {F.~W.}\ \bibnamefont {Hekking}}, \ and\
  \bibinfo {author} {\bibfnamefont {A.}~\bibnamefont {Minguzzi}},\ }\href@noop
  {} {\bibfield  {journal} {\bibinfo  {journal} {Phys. Rev. Lett.}\ }\textbf
  {\bibinfo {volume} {121}},\ \bibinfo {pages} {090404} (\bibinfo {year}
  {2018})}\BibitemShut {NoStop}%
\bibitem [{\citenamefont {Wang}\ \emph {et~al.}(2005)\citenamefont {Wang},
  \citenamefont {Anderson}, \citenamefont {Bright}, \citenamefont {Cornell},
  \citenamefont {Diot}, \citenamefont {Kishimoto}, \citenamefont {Prentiss},
  \citenamefont {Saravanan}, \citenamefont {Segal},\ and\ \citenamefont
  {Wu}}]{wang2005atom}%
  \BibitemOpen
  \bibfield  {author} {\bibinfo {author} {\bibfnamefont {Y.-J.}\ \bibnamefont
  {Wang}}, \bibinfo {author} {\bibfnamefont {D.~Z.}\ \bibnamefont {Anderson}},
  \bibinfo {author} {\bibfnamefont {V.~M.}\ \bibnamefont {Bright}}, \bibinfo
  {author} {\bibfnamefont {E.~A.}\ \bibnamefont {Cornell}}, \bibinfo {author}
  {\bibfnamefont {Q.}~\bibnamefont {Diot}}, \bibinfo {author} {\bibfnamefont
  {T.}~\bibnamefont {Kishimoto}}, \bibinfo {author} {\bibfnamefont
  {M.}~\bibnamefont {Prentiss}}, \bibinfo {author} {\bibfnamefont
  {R.}~\bibnamefont {Saravanan}}, \bibinfo {author} {\bibfnamefont {S.~R.}\
  \bibnamefont {Segal}}, \ and\ \bibinfo {author} {\bibfnamefont
  {S.}~\bibnamefont {Wu}},\ }\href@noop {} {\bibfield  {journal} {\bibinfo
  {journal} {Phys. Rev. Lett.}\ }\textbf {\bibinfo {volume} {94}},\ \bibinfo
  {pages} {090405} (\bibinfo {year} {2005})}\BibitemShut {NoStop}%
\bibitem [{\citenamefont {Schumm}\ \emph {et~al.}(2005)\citenamefont {Schumm},
  \citenamefont {Hofferberth}, \citenamefont {Andersson}, \citenamefont
  {Wildermuth}, \citenamefont {Groth}, \citenamefont {Bar-Joseph},
  \citenamefont {Schmiedmayer},\ and\ \citenamefont
  {Kr{\"u}ger}}]{schumm2005matter}%
  \BibitemOpen
  \bibfield  {author} {\bibinfo {author} {\bibfnamefont {T.}~\bibnamefont
  {Schumm}}, \bibinfo {author} {\bibfnamefont {S.}~\bibnamefont {Hofferberth}},
  \bibinfo {author} {\bibfnamefont {L.~M.}\ \bibnamefont {Andersson}}, \bibinfo
  {author} {\bibfnamefont {S.}~\bibnamefont {Wildermuth}}, \bibinfo {author}
  {\bibfnamefont {S.}~\bibnamefont {Groth}}, \bibinfo {author} {\bibfnamefont
  {I.}~\bibnamefont {Bar-Joseph}}, \bibinfo {author} {\bibfnamefont
  {J.}~\bibnamefont {Schmiedmayer}}, \ and\ \bibinfo {author} {\bibfnamefont
  {P.}~\bibnamefont {Kr{\"u}ger}},\ }\href@noop {} {\bibfield  {journal}
  {\bibinfo  {journal} {Nat. Phys.}\ }\textbf {\bibinfo {volume} {1}},\
  \bibinfo {pages} {57} (\bibinfo {year} {2005})}\BibitemShut {NoStop}%
\bibitem [{\citenamefont {Pezz{\`e}}\ \emph {et~al.}(2018)\citenamefont
  {Pezz{\`e}}, \citenamefont {Smerzi}, \citenamefont {Oberthaler},
  \citenamefont {Schmied},\ and\ \citenamefont {Treutlein}}]{pezze2018quantum}%
  \BibitemOpen
  \bibfield  {author} {\bibinfo {author} {\bibfnamefont {L.}~\bibnamefont
  {Pezz{\`e}}}, \bibinfo {author} {\bibfnamefont {A.}~\bibnamefont {Smerzi}},
  \bibinfo {author} {\bibfnamefont {M.~K.}\ \bibnamefont {Oberthaler}},
  \bibinfo {author} {\bibfnamefont {R.}~\bibnamefont {Schmied}}, \ and\
  \bibinfo {author} {\bibfnamefont {P.}~\bibnamefont {Treutlein}},\ }\href@noop
  {} {\bibfield  {journal} {\bibinfo  {journal} {Rev. Mod. Phys.}\ }\textbf
  {\bibinfo {volume} {90}},\ \bibinfo {pages} {035005} (\bibinfo {year}
  {2018})}\BibitemShut {NoStop}%
\bibitem [{\citenamefont {Zhai}(2015)}]{zhai2015degenerate}%
  \BibitemOpen
  \bibfield  {author} {\bibinfo {author} {\bibfnamefont {H.}~\bibnamefont
  {Zhai}},\ }\href@noop {} {\bibfield  {journal} {\bibinfo  {journal} {Rep.
  Prog. Phys.}\ }\textbf {\bibinfo {volume} {78}},\ \bibinfo {pages} {026001}
  (\bibinfo {year} {2015})}\BibitemShut {NoStop}%
\bibitem [{\citenamefont {Dresselhaus}(1955)}]{dresselhaus1955spin}%
  \BibitemOpen
  \bibfield  {author} {\bibinfo {author} {\bibfnamefont {G.}~\bibnamefont
  {Dresselhaus}},\ }\href@noop {} {\bibfield  {journal} {\bibinfo  {journal}
  {Phys. Rev.}\ }\textbf {\bibinfo {volume} {100}},\ \bibinfo {pages} {580}
  (\bibinfo {year} {1955})}\BibitemShut {NoStop}%
\bibitem [{\citenamefont {Bychkov}\ and\ \citenamefont
  {Rashba}(1984)}]{bychkov1984oscillatory}%
  \BibitemOpen
  \bibfield  {author} {\bibinfo {author} {\bibfnamefont {Y.~A.}\ \bibnamefont
  {Bychkov}}\ and\ \bibinfo {author} {\bibfnamefont {E.~I.}\ \bibnamefont
  {Rashba}},\ }\href@noop {} {\bibfield  {journal} {\bibinfo  {journal} {J.
  Phys. C: Solid State Phys.}\ }\textbf {\bibinfo {volume} {17}},\ \bibinfo
  {pages} {6039} (\bibinfo {year} {1984})}\BibitemShut {NoStop}%
\bibitem [{\citenamefont {Liu}\ \emph {et~al.}(2009)\citenamefont {Liu},
  \citenamefont {Borunda}, \citenamefont {Liu},\ and\ \citenamefont
  {Sinova}}]{liu_effect_2009}%
  \BibitemOpen
  \bibfield  {author} {\bibinfo {author} {\bibfnamefont {X.-J.}\ \bibnamefont
  {Liu}}, \bibinfo {author} {\bibfnamefont {M.~F.}\ \bibnamefont {Borunda}},
  \bibinfo {author} {\bibfnamefont {X.}~\bibnamefont {Liu}}, \ and\ \bibinfo
  {author} {\bibfnamefont {J.}~\bibnamefont {Sinova}},\ }\href {\doibase
  10.1103/PhysRevLett.102.046402} {\bibfield  {journal} {\bibinfo  {journal}
  {Phys. Rev. Lett.}\ }\textbf {\bibinfo {volume} {102}},\ \bibinfo {pages}
  {046402} (\bibinfo {year} {2009})}\BibitemShut {NoStop}%
\bibitem [{\citenamefont {Lin}\ \emph {et~al.}(2009)\citenamefont {Lin},
  \citenamefont {Compton}, \citenamefont {Perry}, \citenamefont {Phillips},
  \citenamefont {Porto},\ and\ \citenamefont {Spielman}}]{lin2009bose}%
  \BibitemOpen
  \bibfield  {author} {\bibinfo {author} {\bibfnamefont {Y.-J.}\ \bibnamefont
  {Lin}}, \bibinfo {author} {\bibfnamefont {R.~L.}\ \bibnamefont {Compton}},
  \bibinfo {author} {\bibfnamefont {A.~R.}\ \bibnamefont {Perry}}, \bibinfo
  {author} {\bibfnamefont {W.~D.}\ \bibnamefont {Phillips}}, \bibinfo {author}
  {\bibfnamefont {J.~V.}\ \bibnamefont {Porto}}, \ and\ \bibinfo {author}
  {\bibfnamefont {I.~B.}\ \bibnamefont {Spielman}},\ }\href@noop {} {\bibfield
  {journal} {\bibinfo  {journal} {Phys. Rev. Lett.}\ }\textbf {\bibinfo
  {volume} {102}},\ \bibinfo {pages} {130401} (\bibinfo {year}
  {2009})}\BibitemShut {NoStop}%
\bibitem [{\citenamefont {Chunji}\ \emph {et~al.}(2010)\citenamefont {Chunji},
  \citenamefont {Chao}, \citenamefont {Chaoming},\ and\ \citenamefont
  {Hui}}]{chunji2010spin}%
  \BibitemOpen
  \bibfield  {author} {\bibinfo {author} {\bibfnamefont {W.}~\bibnamefont
  {Chunji}}, \bibinfo {author} {\bibfnamefont {G.}~\bibnamefont {Chao}},
  \bibinfo {author} {\bibfnamefont {J.}~\bibnamefont {Chaoming}}, \ and\
  \bibinfo {author} {\bibfnamefont {Z.}~\bibnamefont {Hui}},\ }\href@noop {}
  {\bibfield  {journal} {\bibinfo  {journal} {Phys. Rev. Lett.}\ }\textbf
  {\bibinfo {volume} {105}} (\bibinfo {year} {2010})}\BibitemShut {NoStop}%
\bibitem [{\citenamefont {Lin}\ \emph {et~al.}(2011)\citenamefont {Lin},
  \citenamefont {Jimenez-Garcia},\ and\ \citenamefont
  {Spielman}}]{lin2011spin}%
  \BibitemOpen
  \bibfield  {author} {\bibinfo {author} {\bibfnamefont {Y.-J.}\ \bibnamefont
  {Lin}}, \bibinfo {author} {\bibfnamefont {K.}~\bibnamefont {Jimenez-Garcia}},
  \ and\ \bibinfo {author} {\bibfnamefont {I.~B.}\ \bibnamefont {Spielman}},\
  }\href@noop {} {\bibfield  {journal} {\bibinfo  {journal} {Nature}\ }\textbf
  {\bibinfo {volume} {471}},\ \bibinfo {pages} {83} (\bibinfo {year}
  {2011})}\BibitemShut {NoStop}%
\bibitem [{\citenamefont {Li}\ \emph {et~al.}(2013)\citenamefont {Li},
  \citenamefont {Martone}, \citenamefont {Pitaevskii},\ and\ \citenamefont
  {Stringari}}]{li2013superstripes}%
  \BibitemOpen
  \bibfield  {author} {\bibinfo {author} {\bibfnamefont {Y.}~\bibnamefont
  {Li}}, \bibinfo {author} {\bibfnamefont {G.~I.}\ \bibnamefont {Martone}},
  \bibinfo {author} {\bibfnamefont {L.~P.}\ \bibnamefont {Pitaevskii}}, \ and\
  \bibinfo {author} {\bibfnamefont {S.}~\bibnamefont {Stringari}},\ }\href@noop
  {} {\bibfield  {journal} {\bibinfo  {journal} {Phys. Rev. Lett.}\ }\textbf
  {\bibinfo {volume} {110}},\ \bibinfo {pages} {235302} (\bibinfo {year}
  {2013})}\BibitemShut {NoStop}%
\bibitem [{\citenamefont {Bao}\ and\ \citenamefont
  {Cai}(2015)}]{bao2015ground}%
  \BibitemOpen
  \bibfield  {author} {\bibinfo {author} {\bibfnamefont {W.}~\bibnamefont
  {Bao}}\ and\ \bibinfo {author} {\bibfnamefont {Y.}~\bibnamefont {Cai}},\
  }\href@noop {} {\bibfield  {journal} {\bibinfo  {journal} {SIAM J. Appl.
  Math.}\ }\textbf {\bibinfo {volume} {75}},\ \bibinfo {pages} {492} (\bibinfo
  {year} {2015})}\BibitemShut {NoStop}%
\bibitem [{\citenamefont {Bhuvaneswari}\ \emph {et~al.}(2018)\citenamefont
  {Bhuvaneswari}, \citenamefont {Nithyanandan},\ and\ \citenamefont
  {Muruganandam}}]{bhuvaneswari2018spotlighting}%
  \BibitemOpen
  \bibfield  {author} {\bibinfo {author} {\bibfnamefont {S.}~\bibnamefont
  {Bhuvaneswari}}, \bibinfo {author} {\bibfnamefont {K.}~\bibnamefont
  {Nithyanandan}}, \ and\ \bibinfo {author} {\bibfnamefont {P.}~\bibnamefont
  {Muruganandam}},\ }\href@noop {} {\bibfield  {journal} {\bibinfo  {journal}
  {J. Phys. Commun.}\ }\textbf {\bibinfo {volume} {2}},\ \bibinfo {pages}
  {025008} (\bibinfo {year} {2018})}\BibitemShut {NoStop}%
\bibitem [{\citenamefont {Jochim}\ \emph {et~al.}(2003)\citenamefont {Jochim},
  \citenamefont {Bartenstein}, \citenamefont {Altmeyer}, \citenamefont {Hendl},
  \citenamefont {Riedl}, \citenamefont {Chin}, \citenamefont {Denschlag},\ and\
  \citenamefont {Grimm}}]{jochim2003bose}%
  \BibitemOpen
  \bibfield  {author} {\bibinfo {author} {\bibfnamefont {S.}~\bibnamefont
  {Jochim}}, \bibinfo {author} {\bibfnamefont {M.}~\bibnamefont {Bartenstein}},
  \bibinfo {author} {\bibfnamefont {A.}~\bibnamefont {Altmeyer}}, \bibinfo
  {author} {\bibfnamefont {G.}~\bibnamefont {Hendl}}, \bibinfo {author}
  {\bibfnamefont {S.}~\bibnamefont {Riedl}}, \bibinfo {author} {\bibfnamefont
  {C.}~\bibnamefont {Chin}}, \bibinfo {author} {\bibfnamefont {J.~H.}\
  \bibnamefont {Denschlag}}, \ and\ \bibinfo {author} {\bibfnamefont
  {R.}~\bibnamefont {Grimm}},\ }\href@noop {} {\bibfield  {journal} {\bibinfo
  {journal} {Science}\ }\textbf {\bibinfo {volume} {302}},\ \bibinfo {pages}
  {2101} (\bibinfo {year} {2003})}\BibitemShut {NoStop}%
\bibitem [{\citenamefont {Bradley}\ \emph {et~al.}(1995)\citenamefont
  {Bradley}, \citenamefont {Sackett}, \citenamefont {Tollett},\ and\
  \citenamefont {Hulet}}]{bradley1995evidence}%
  \BibitemOpen
  \bibfield  {author} {\bibinfo {author} {\bibfnamefont {C.~C.}\ \bibnamefont
  {Bradley}}, \bibinfo {author} {\bibfnamefont {C.}~\bibnamefont {Sackett}},
  \bibinfo {author} {\bibfnamefont {J.}~\bibnamefont {Tollett}}, \ and\
  \bibinfo {author} {\bibfnamefont {R.~G.}\ \bibnamefont {Hulet}},\ }\href@noop
  {} {\bibfield  {journal} {\bibinfo  {journal} {Phys. Rev. Lett.}\ }\textbf
  {\bibinfo {volume} {75}},\ \bibinfo {pages} {1687} (\bibinfo {year}
  {1995})}\BibitemShut {NoStop}%
\bibitem [{\citenamefont {Wang}\ \emph {et~al.}(2010)\citenamefont {Wang},
  \citenamefont {Gao}, \citenamefont {Jian},\ and\ \citenamefont
  {Zhai}}]{wang2010spin}%
  \BibitemOpen
  \bibfield  {author} {\bibinfo {author} {\bibfnamefont {C.}~\bibnamefont
  {Wang}}, \bibinfo {author} {\bibfnamefont {C.}~\bibnamefont {Gao}}, \bibinfo
  {author} {\bibfnamefont {C.-M.}\ \bibnamefont {Jian}}, \ and\ \bibinfo
  {author} {\bibfnamefont {H.}~\bibnamefont {Zhai}},\ }\href@noop {} {\bibfield
   {journal} {\bibinfo  {journal} {Phys. Rev. Lett.}\ }\textbf {\bibinfo
  {volume} {105}},\ \bibinfo {pages} {160403} (\bibinfo {year}
  {2010})}\BibitemShut {NoStop}%
\bibitem [{\citenamefont {Zhai}(2012)}]{zhai2012spin}%
  \BibitemOpen
  \bibfield  {author} {\bibinfo {author} {\bibfnamefont {H.}~\bibnamefont
  {Zhai}},\ }\href@noop {} {\bibfield  {journal} {\bibinfo  {journal} {Int. J.
  Mod. Phys. B}\ }\textbf {\bibinfo {volume} {26}},\ \bibinfo {pages} {1230001}
  (\bibinfo {year} {2012})}\BibitemShut {NoStop}%
\bibitem [{\citenamefont {Zheng}\ \emph {et~al.}(2013)\citenamefont {Zheng},
  \citenamefont {Yu}, \citenamefont {Cui},\ and\ \citenamefont
  {Zhai}}]{zheng2013properties}%
  \BibitemOpen
  \bibfield  {author} {\bibinfo {author} {\bibfnamefont {W.}~\bibnamefont
  {Zheng}}, \bibinfo {author} {\bibfnamefont {Z.-Q.}\ \bibnamefont {Yu}},
  \bibinfo {author} {\bibfnamefont {X.}~\bibnamefont {Cui}}, \ and\ \bibinfo
  {author} {\bibfnamefont {H.}~\bibnamefont {Zhai}},\ }\href@noop {} {\bibfield
   {journal} {\bibinfo  {journal} {J. Phys. B: At. Mol. Opt. Phys.}\ }\textbf
  {\bibinfo {volume} {46}},\ \bibinfo {pages} {134007} (\bibinfo {year}
  {2013})}\BibitemShut {NoStop}%
\bibitem [{\citenamefont {Ramachandhran}\ \emph {et~al.}(2012)\citenamefont
  {Ramachandhran}, \citenamefont {Opanchuk}, \citenamefont {Liu}, \citenamefont
  {Pu}, \citenamefont {Drummond},\ and\ \citenamefont
  {Hu}}]{ramachandhran2012half}%
  \BibitemOpen
  \bibfield  {author} {\bibinfo {author} {\bibfnamefont {B.}~\bibnamefont
  {Ramachandhran}}, \bibinfo {author} {\bibfnamefont {B.}~\bibnamefont
  {Opanchuk}}, \bibinfo {author} {\bibfnamefont {X.-J.}\ \bibnamefont {Liu}},
  \bibinfo {author} {\bibfnamefont {H.}~\bibnamefont {Pu}}, \bibinfo {author}
  {\bibfnamefont {P.~D.}\ \bibnamefont {Drummond}}, \ and\ \bibinfo {author}
  {\bibfnamefont {H.}~\bibnamefont {Hu}},\ }\href@noop {} {\bibfield  {journal}
  {\bibinfo  {journal} {Phys. Rev. A}\ }\textbf {\bibinfo {volume} {85}},\
  \bibinfo {pages} {023606} (\bibinfo {year} {2012})}\BibitemShut {NoStop}%
\bibitem [{\citenamefont {Hu}\ \emph {et~al.}(2012)\citenamefont {Hu},
  \citenamefont {Ramachandhran}, \citenamefont {Pu},\ and\ \citenamefont
  {Liu}}]{hu2012spin}%
  \BibitemOpen
  \bibfield  {author} {\bibinfo {author} {\bibfnamefont {H.}~\bibnamefont
  {Hu}}, \bibinfo {author} {\bibfnamefont {B.}~\bibnamefont {Ramachandhran}},
  \bibinfo {author} {\bibfnamefont {H.}~\bibnamefont {Pu}}, \ and\ \bibinfo
  {author} {\bibfnamefont {X.-J.}\ \bibnamefont {Liu}},\ }\href@noop {}
  {\bibfield  {journal} {\bibinfo  {journal} {Phys. Rev. Lett.}\ }\textbf
  {\bibinfo {volume} {108}},\ \bibinfo {pages} {010402} (\bibinfo {year}
  {2012})}\BibitemShut {NoStop}%
\bibitem [{\citenamefont {Cong-Jun}\ \emph {et~al.}(2011)\citenamefont
  {Cong-Jun}, \citenamefont {Mondragon-Shem},\ and\ \citenamefont
  {Xiang-Fa}}]{cong2011unconventional}%
  \BibitemOpen
  \bibfield  {author} {\bibinfo {author} {\bibfnamefont {W.}~\bibnamefont
  {Cong-Jun}}, \bibinfo {author} {\bibfnamefont {I.}~\bibnamefont
  {Mondragon-Shem}}, \ and\ \bibinfo {author} {\bibfnamefont {Z.}~\bibnamefont
  {Xiang-Fa}},\ }\href@noop {} {\bibfield  {journal} {\bibinfo  {journal}
  {Chin. Phys. Lett.}\ }\textbf {\bibinfo {volume} {28}},\ \bibinfo {pages}
  {097102} (\bibinfo {year} {2011})}\BibitemShut {NoStop}%
\bibitem [{\citenamefont {Milburn}\ \emph {et~al.}(1997)\citenamefont
  {Milburn}, \citenamefont {Corney}, \citenamefont {Wright},\ and\
  \citenamefont {Walls}}]{milburn1997quantum}%
  \BibitemOpen
  \bibfield  {author} {\bibinfo {author} {\bibfnamefont {G.}~\bibnamefont
  {Milburn}}, \bibinfo {author} {\bibfnamefont {J.}~\bibnamefont {Corney}},
  \bibinfo {author} {\bibfnamefont {E.~M.}\ \bibnamefont {Wright}}, \ and\
  \bibinfo {author} {\bibfnamefont {D.}~\bibnamefont {Walls}},\ }\href@noop {}
  {\bibfield  {journal} {\bibinfo  {journal} {Phys. Rev. A}\ }\textbf {\bibinfo
  {volume} {55}},\ \bibinfo {pages} {4318} (\bibinfo {year}
  {1997})}\BibitemShut {NoStop}%
\bibitem [{\citenamefont {Shin}\ \emph {et~al.}(2004)\citenamefont {Shin},
  \citenamefont {Saba}, \citenamefont {Pasquini}, \citenamefont {Ketterle},
  \citenamefont {Pritchard},\ and\ \citenamefont {Leanhardt}}]{shin2004atom}%
  \BibitemOpen
  \bibfield  {author} {\bibinfo {author} {\bibfnamefont {Y.}~\bibnamefont
  {Shin}}, \bibinfo {author} {\bibfnamefont {M.}~\bibnamefont {Saba}}, \bibinfo
  {author} {\bibfnamefont {T.}~\bibnamefont {Pasquini}}, \bibinfo {author}
  {\bibfnamefont {W.}~\bibnamefont {Ketterle}}, \bibinfo {author}
  {\bibfnamefont {D.}~\bibnamefont {Pritchard}}, \ and\ \bibinfo {author}
  {\bibfnamefont {A.}~\bibnamefont {Leanhardt}},\ }\href@noop {} {\bibfield
  {journal} {\bibinfo  {journal} {Phys. Rev. Lett.}\ }\textbf {\bibinfo
  {volume} {92}},\ \bibinfo {pages} {050405} (\bibinfo {year}
  {2004})}\BibitemShut {NoStop}%
\bibitem [{\citenamefont {M{\"u}stecapl{\i}o{\u{g}}lu}\ \emph
  {et~al.}(2007)\citenamefont {M{\"u}stecapl{\i}o{\u{g}}lu}, \citenamefont
  {Zhang},\ and\ \citenamefont {You}}]{mustecapliouglu2007quantum}%
  \BibitemOpen
  \bibfield  {author} {\bibinfo {author} {\bibfnamefont {{\"O}.}~\bibnamefont
  {M{\"u}stecapl{\i}o{\u{g}}lu}}, \bibinfo {author} {\bibfnamefont
  {W.}~\bibnamefont {Zhang}}, \ and\ \bibinfo {author} {\bibfnamefont
  {L.}~\bibnamefont {You}},\ }\href@noop {} {\bibfield  {journal} {\bibinfo
  {journal} {Phys. Rev. A}\ }\textbf {\bibinfo {volume} {75}},\ \bibinfo
  {pages} {023605} (\bibinfo {year} {2007})}\BibitemShut {NoStop}%
\bibitem [{\citenamefont {Jin}\ \emph {et~al.}(1996)\citenamefont {Jin},
  \citenamefont {Ensher}, \citenamefont {Matthews}, \citenamefont {Wieman},\
  and\ \citenamefont {Cornell}}]{jin1996collective}%
  \BibitemOpen
  \bibfield  {author} {\bibinfo {author} {\bibfnamefont {D.}~\bibnamefont
  {Jin}}, \bibinfo {author} {\bibfnamefont {J.}~\bibnamefont {Ensher}},
  \bibinfo {author} {\bibfnamefont {M.}~\bibnamefont {Matthews}}, \bibinfo
  {author} {\bibfnamefont {C.}~\bibnamefont {Wieman}}, \ and\ \bibinfo {author}
  {\bibfnamefont {E.~A.}\ \bibnamefont {Cornell}},\ }\href@noop {} {\bibfield
  {journal} {\bibinfo  {journal} {Phys. Rev. Lett.}\ }\textbf {\bibinfo
  {volume} {77}},\ \bibinfo {pages} {420} (\bibinfo {year} {1996})}\BibitemShut
  {NoStop}%
\bibitem [{\citenamefont {Mewes}\ \emph
  {et~al.}(1996{\natexlab{a}})\citenamefont {Mewes}, \citenamefont {Andrews},
  \citenamefont {Van~Druten}, \citenamefont {Kurn}, \citenamefont {Durfee},
  \citenamefont {Townsend},\ and\ \citenamefont
  {Ketterle}}]{mewes1996collective}%
  \BibitemOpen
  \bibfield  {author} {\bibinfo {author} {\bibfnamefont {M.-O.}\ \bibnamefont
  {Mewes}}, \bibinfo {author} {\bibfnamefont {M.}~\bibnamefont {Andrews}},
  \bibinfo {author} {\bibfnamefont {N.}~\bibnamefont {Van~Druten}}, \bibinfo
  {author} {\bibfnamefont {D.}~\bibnamefont {Kurn}}, \bibinfo {author}
  {\bibfnamefont {D.}~\bibnamefont {Durfee}}, \bibinfo {author} {\bibfnamefont
  {C.}~\bibnamefont {Townsend}}, \ and\ \bibinfo {author} {\bibfnamefont
  {W.}~\bibnamefont {Ketterle}},\ }\href@noop {} {\bibfield  {journal}
  {\bibinfo  {journal} {Phys. Rev. Lett.}\ }\textbf {\bibinfo {volume} {77}},\
  \bibinfo {pages} {988} (\bibinfo {year} {1996}{\natexlab{a}})}\BibitemShut
  {NoStop}%
\bibitem [{\citenamefont {Edwards}\ \emph {et~al.}(1996)\citenamefont
  {Edwards}, \citenamefont {Ruprecht}, \citenamefont {Burnett}, \citenamefont
  {Dodd},\ and\ \citenamefont {Clark}}]{edwards1996collective}%
  \BibitemOpen
  \bibfield  {author} {\bibinfo {author} {\bibfnamefont {M.}~\bibnamefont
  {Edwards}}, \bibinfo {author} {\bibfnamefont {P.}~\bibnamefont {Ruprecht}},
  \bibinfo {author} {\bibfnamefont {K.}~\bibnamefont {Burnett}}, \bibinfo
  {author} {\bibfnamefont {R.}~\bibnamefont {Dodd}}, \ and\ \bibinfo {author}
  {\bibfnamefont {C.~W.}\ \bibnamefont {Clark}},\ }\href@noop {} {\bibfield
  {journal} {\bibinfo  {journal} {Phys. Rev. Lett.}\ }\textbf {\bibinfo
  {volume} {77}},\ \bibinfo {pages} {1671} (\bibinfo {year}
  {1996})}\BibitemShut {NoStop}%
\bibitem [{\citenamefont {Mewes}\ \emph
  {et~al.}(1996{\natexlab{b}})\citenamefont {Mewes}, \citenamefont {Andrews},
  \citenamefont {Van~Druten}, \citenamefont {Kurn}, \citenamefont {Durfee},\
  and\ \citenamefont {Ketterle}}]{mewes1996bose}%
  \BibitemOpen
  \bibfield  {author} {\bibinfo {author} {\bibfnamefont {M.-O.}\ \bibnamefont
  {Mewes}}, \bibinfo {author} {\bibfnamefont {M.}~\bibnamefont {Andrews}},
  \bibinfo {author} {\bibfnamefont {N.}~\bibnamefont {Van~Druten}}, \bibinfo
  {author} {\bibfnamefont {D.}~\bibnamefont {Kurn}}, \bibinfo {author}
  {\bibfnamefont {D.}~\bibnamefont {Durfee}}, \ and\ \bibinfo {author}
  {\bibfnamefont {W.}~\bibnamefont {Ketterle}},\ }\href@noop {} {\bibfield
  {journal} {\bibinfo  {journal} {Phys. Rev. Lett.}\ }\textbf {\bibinfo
  {volume} {77}},\ \bibinfo {pages} {416} (\bibinfo {year}
  {1996}{\natexlab{b}})}\BibitemShut {NoStop}%
\bibitem [{\citenamefont {Rauer}\ \emph {et~al.}(2016)\citenamefont {Rauer},
  \citenamefont {Gri{\v{s}}ins}, \citenamefont {Mazets}, \citenamefont
  {Schweigler}, \citenamefont {Rohringer}, \citenamefont {Geiger},
  \citenamefont {Langen},\ and\ \citenamefont
  {Schmiedmayer}}]{rauer2016cooling}%
  \BibitemOpen
  \bibfield  {author} {\bibinfo {author} {\bibfnamefont {B.}~\bibnamefont
  {Rauer}}, \bibinfo {author} {\bibfnamefont {P.}~\bibnamefont
  {Gri{\v{s}}ins}}, \bibinfo {author} {\bibfnamefont {I.~E.}\ \bibnamefont
  {Mazets}}, \bibinfo {author} {\bibfnamefont {T.}~\bibnamefont {Schweigler}},
  \bibinfo {author} {\bibfnamefont {W.}~\bibnamefont {Rohringer}}, \bibinfo
  {author} {\bibfnamefont {R.}~\bibnamefont {Geiger}}, \bibinfo {author}
  {\bibfnamefont {T.}~\bibnamefont {Langen}}, \ and\ \bibinfo {author}
  {\bibfnamefont {J.}~\bibnamefont {Schmiedmayer}},\ }\href@noop {} {\bibfield
  {journal} {\bibinfo  {journal} {Phys. Rev. Lett.}\ }\textbf {\bibinfo
  {volume} {116}},\ \bibinfo {pages} {030402} (\bibinfo {year}
  {2016})}\BibitemShut {NoStop}%
\bibitem [{\citenamefont {Pigneur}\ \emph {et~al.}(2018)\citenamefont
  {Pigneur}, \citenamefont {Berrada}, \citenamefont {Bonneau}, \citenamefont
  {Schumm}, \citenamefont {Demler},\ and\ \citenamefont
  {Schmiedmayer}}]{pigneur2018relaxation}%
  \BibitemOpen
  \bibfield  {author} {\bibinfo {author} {\bibfnamefont {M.}~\bibnamefont
  {Pigneur}}, \bibinfo {author} {\bibfnamefont {T.}~\bibnamefont {Berrada}},
  \bibinfo {author} {\bibfnamefont {M.}~\bibnamefont {Bonneau}}, \bibinfo
  {author} {\bibfnamefont {T.}~\bibnamefont {Schumm}}, \bibinfo {author}
  {\bibfnamefont {E.}~\bibnamefont {Demler}}, \ and\ \bibinfo {author}
  {\bibfnamefont {J.}~\bibnamefont {Schmiedmayer}},\ }\href@noop {} {\bibfield
  {journal} {\bibinfo  {journal} {Phys. Rev. Lett.}\ }\textbf {\bibinfo
  {volume} {120}},\ \bibinfo {pages} {173601} (\bibinfo {year}
  {2018})}\BibitemShut {NoStop}%
\bibitem [{V1V()}]{V1V2V12}%
  \BibitemOpen
  \href@noop {} {}\bibinfo {note} {The integrals for these three parameters
  are: $V_{m} = \int
  g\big(|\varphi_{m-2\uparrow}|^{4}+|\varphi_{m-2\downarrow}|^{4}+2c_{12}|\varphi_{m-2\uparrow}|^{2}|\varphi_{m-2\downarrow}|^{2}\big)d\bm{r}$
  for $m= 1$, $2$, and $V_{12} = \frac{1}{2} \int
  g\big(2|\varphi_{0\uparrow}|^{2}|\varphi_{-1\uparrow}|^{2}+2|\varphi_{-1\downarrow}|^{2}|\varphi_{0\downarrow}|^{2}+
  c_{12}|\varphi_{-1\uparrow}|^{2}|\varphi_{0\downarrow}|^{2}+c_{12}|\varphi_{-1\downarrow}|^{2}|\varphi_{0\uparrow}|^{2}+
  c_{12}\varphi_{-1\uparrow}^{\ast}\varphi_{-1\downarrow}^{\ast}\varphi_{0\downarrow}\varphi_{0\uparrow}+\text{h.c.}\big)d\bm{r}
  $.}\BibitemShut {Stop}%
\bibitem [{\citenamefont {Bloch}(2005)}]{bloch2005ultracold}%
  \BibitemOpen
  \bibfield  {author} {\bibinfo {author} {\bibfnamefont {I.}~\bibnamefont
  {Bloch}},\ }\href@noop {} {\bibfield  {journal} {\bibinfo  {journal} {Nat.
  Phys.}\ }\textbf {\bibinfo {volume} {1}},\ \bibinfo {pages} {23} (\bibinfo
  {year} {2005})}\BibitemShut {NoStop}%
\bibitem [{\citenamefont {Liu}\ \emph {et~al.}(2007)\citenamefont {Liu},
  \citenamefont {Fu}, \citenamefont {Yang},\ and\ \citenamefont
  {Liu}}]{liu2007josephson}%
  \BibitemOpen
  \bibfield  {author} {\bibinfo {author} {\bibfnamefont {B.}~\bibnamefont
  {Liu}}, \bibinfo {author} {\bibfnamefont {L.-B.}\ \bibnamefont {Fu}},
  \bibinfo {author} {\bibfnamefont {S.-P.}\ \bibnamefont {Yang}}, \ and\
  \bibinfo {author} {\bibfnamefont {J.}~\bibnamefont {Liu}},\ }\href@noop {}
  {\bibfield  {journal} {\bibinfo  {journal} {Phys. Rev. A}\ }\textbf {\bibinfo
  {volume} {75}},\ \bibinfo {pages} {033601} (\bibinfo {year}
  {2007})}\BibitemShut {NoStop}%
\bibitem [{\citenamefont {Trombettoni}\ and\ \citenamefont
  {Smerzi}(2001)}]{trombettoni2001discrete}%
  \BibitemOpen
  \bibfield  {author} {\bibinfo {author} {\bibfnamefont {A.}~\bibnamefont
  {Trombettoni}}\ and\ \bibinfo {author} {\bibfnamefont {A.}~\bibnamefont
  {Smerzi}},\ }\href@noop {} {\bibfield  {journal} {\bibinfo  {journal} {Phys.
  Rev. Lett.}\ }\textbf {\bibinfo {volume} {86}},\ \bibinfo {pages} {2353}
  (\bibinfo {year} {2001})}\BibitemShut {NoStop}%
\bibitem [{DCJ()}]{DCJ}%
  \BibitemOpen
  \href@noop {} {}\bibinfo {note} {In the Josepshson junction considered by
  Feynman \cite{feynman1965lectures}, the current and phase are described by $I
  = \dot{z} = K \sin(\theta)$ and $\dot{\theta} = (\mu_0 + A \cos(\omega t))$.
  From the second equation, $\theta(t) = \theta_0 + \mu_0 t + A \sin(\omega
  t)/\omega$. The current $I$ in this case has exactly the same form as $z(t)$
  considered in Eq. \ref{eq-app}.}\BibitemShut {Stop}%
\bibitem [{str()}]{strongomega2}%
  \BibitemOpen
  \href@noop {} {}\bibinfo {note} {In the strong modulating limit when
  $\omega_2 > \omega$, the excited bands may also be excited. We find that when
  $\omega_2 = 0.8\omega$ ($\eta = 80\%$), the excitation of the states are
  still negligible for the parameters used in Fig. \ref{fig-fig4}
  (b).}\BibitemShut {Stop}%
\bibitem [{est()}]{estimateTmillsecond}%
  \BibitemOpen
  \href@noop {} {}\bibinfo {note} {Using the transverse confinement frequency
  in one dimensional BEC \cite{rauer2016cooling,pigneur2018relaxation} with
  $\omega = 2\pi \cdot 2.0$ kHz, $\alpha_J = 0.9$, $h_x = 2\pi\cdot 0.8$ kHz
  ($\eta = 40\%$), we estimate $T = 1.4$ ms.}\BibitemShut {Stop}%
\bibitem [{\citenamefont {Pan}\ and\ \citenamefont
  {Draayer}(1999)}]{pan1999analytical}%
  \BibitemOpen
  \bibfield  {author} {\bibinfo {author} {\bibfnamefont {F.}~\bibnamefont
  {Pan}}\ and\ \bibinfo {author} {\bibfnamefont {J.}~\bibnamefont {Draayer}},\
  }\href@noop {} {\bibfield  {journal} {\bibinfo  {journal} {Phys. Lett. B}\
  }\textbf {\bibinfo {volume} {451}},\ \bibinfo {pages} {1} (\bibinfo {year}
  {1999})}\BibitemShut {NoStop}%
\bibitem [{\citenamefont {Ma}\ \emph {et~al.}(2011)\citenamefont {Ma},
  \citenamefont {Wang}, \citenamefont {Sun},\ and\ \citenamefont
  {Nori}}]{ma2011quantum}%
  \BibitemOpen
  \bibfield  {author} {\bibinfo {author} {\bibfnamefont {J.}~\bibnamefont
  {Ma}}, \bibinfo {author} {\bibfnamefont {X.}~\bibnamefont {Wang}}, \bibinfo
  {author} {\bibfnamefont {C.-P.}\ \bibnamefont {Sun}}, \ and\ \bibinfo
  {author} {\bibfnamefont {F.}~\bibnamefont {Nori}},\ }\href@noop {} {\bibfield
   {journal} {\bibinfo  {journal} {Phys. Rep.}\ }\textbf {\bibinfo {volume}
  {509}},\ \bibinfo {pages} {89} (\bibinfo {year} {2011})}\BibitemShut
  {NoStop}%
\bibitem [{\citenamefont {Salvatori}\ \emph {et~al.}(2014)\citenamefont
  {Salvatori}, \citenamefont {Mandarino},\ and\ \citenamefont
  {Paris}}]{salvatori2014quantum}%
  \BibitemOpen
  \bibfield  {author} {\bibinfo {author} {\bibfnamefont {G.}~\bibnamefont
  {Salvatori}}, \bibinfo {author} {\bibfnamefont {A.}~\bibnamefont
  {Mandarino}}, \ and\ \bibinfo {author} {\bibfnamefont {M.~G.}\ \bibnamefont
  {Paris}},\ }\href@noop {} {\bibfield  {journal} {\bibinfo  {journal} {Phys.
  Rev. A}\ }\textbf {\bibinfo {volume} {90}},\ \bibinfo {pages} {022111}
  (\bibinfo {year} {2014})}\BibitemShut {NoStop}%
\bibitem [{\citenamefont {Luo}\ \emph {et~al.}(2017)\citenamefont {Luo},
  \citenamefont {Zou}, \citenamefont {Wu}, \citenamefont {Liu}, \citenamefont
  {Han}, \citenamefont {Tey},\ and\ \citenamefont
  {You}}]{luo2017deterministic}%
  \BibitemOpen
  \bibfield  {author} {\bibinfo {author} {\bibfnamefont {X.-Y.}\ \bibnamefont
  {Luo}}, \bibinfo {author} {\bibfnamefont {Y.-Q.}\ \bibnamefont {Zou}},
  \bibinfo {author} {\bibfnamefont {L.-N.}\ \bibnamefont {Wu}}, \bibinfo
  {author} {\bibfnamefont {Q.}~\bibnamefont {Liu}}, \bibinfo {author}
  {\bibfnamefont {M.-F.}\ \bibnamefont {Han}}, \bibinfo {author} {\bibfnamefont
  {M.~K.}\ \bibnamefont {Tey}}, \ and\ \bibinfo {author} {\bibfnamefont
  {L.}~\bibnamefont {You}},\ }\href@noop {} {\bibfield  {journal} {\bibinfo
  {journal} {Science}\ }\textbf {\bibinfo {volume} {355}},\ \bibinfo {pages}
  {620} (\bibinfo {year} {2017})}\BibitemShut {NoStop}%
\bibitem [{\citenamefont {Graefe}\ and\ \citenamefont
  {Korsch}(2007)}]{graefe2007semiclassical}%
  \BibitemOpen
  \bibfield  {author} {\bibinfo {author} {\bibfnamefont {E.}~\bibnamefont
  {Graefe}}\ and\ \bibinfo {author} {\bibfnamefont {H.}~\bibnamefont
  {Korsch}},\ }\href@noop {} {\bibfield  {journal} {\bibinfo  {journal} {Phys.
  Rev. A}\ }\textbf {\bibinfo {volume} {76}},\ \bibinfo {pages} {032116}
  (\bibinfo {year} {2007})}\BibitemShut {NoStop}%
\end{thebibliography}%

\end{document}